\documentclass[journal]{IEEEtran}
\IEEEoverridecommandlockouts
\usepackage[font={footnotesize},skip=2pt]{caption}
\usepackage[font={footnotesize}]{subcaption}
\usepackage{mathtools}
\usepackage{amsmath}
\usepackage{algpseudocode}
\usepackage[ruled,vlined]{algorithm2e}
\SetKwInput{KwInput}{Input}                % Set the Input
\SetKwInput{KwOutput}{Output}    
\SetKwInput{KwInit}{Initialize} 
\SetKwInput{KWstop}{stop} % set the Output
\usepackage{svg}
\usepackage{amsfonts} 
\usepackage{booktabs} % To thicken table lines
\usepackage{wasysym}
\usepackage{multirow}
\usepackage{multicol}
\usepackage{enumerate}
\usepackage{enumitem}
\usepackage{xcolor}
\usepackage{float}

\usepackage[T1]{fontenc}
\usepackage{anyfontsize}
\usepackage{graphicx}
\usepackage{pgfgantt}
\usepackage{lscape} 
\usepackage{url}
\usepackage{pgfplots}
\pgfplotsset{compat=newest}
\usepackage{adjustbox}

\usepgfplotslibrary{groupplots} % LaTeX and plain TeX
\usetikzlibrary{pgfplots.groupplots} % LaTeX and plain TeX

\usepackage{graphicx}
\usepackage{tikz}
\usetikzlibrary{external}
\usepackage{scalerel}
\usetikzlibrary{svg.path}

\definecolor{orcidlogocol}{HTML}{A6CE39}
\tikzset{
  orcidlogo/.pic={
    \fill[orcidlogocol] svg{M256,128c0,70.7-57.3,128-128,128C57.3,256,0,198.7,0,128C0,57.3,57.3,0,128,0C198.7,0,256,57.3,256,128z};
    \fill[white] svg{M86.3,186.2H70.9V79.1h15.4v48.4V186.2z}
                 svg{M108.9,79.1h41.6c39.6,0,57,28.3,57,53.6c0,27.5-21.5,53.6-56.8,53.6h-41.8V79.1z M124.3,172.4h24.5c34.9,0,42.9-26.5,42.9-39.7c0-21.5-13.7-39.7-43.7-39.7h-23.7V172.4z}
                 svg{M88.7,56.8c0,5.5-4.5,10.1-10.1,10.1c-5.6,0-10.1-4.6-10.1-10.1c0-5.6,4.5-10.1,10.1-10.1C84.2,46.7,88.7,51.3,88.7,56.8z};
  }
}

\newcommand\orcidicon[1]{\href{https://orcid.org/#1}{\mbox{\scalerel*{
\begin{tikzpicture}[yscale=-1,transform shape]
\pic{orcidlogo};
\end{tikzpicture}
}{|}}}}

\usepackage[hidelinks]{hyperref}

\usepackage{xurl}

\usepackage[most]{tcolorbox}
\tcbuselibrary{skins,breakable}

\usepackage{lipsum}
\usepackage{xcolor}

\newtcolorbox{userbubble}{
  colback=black!5,
  colframe=black!40,
  sharp corners=southwest,
  boxrule=0.4pt,
  width=\linewidth,
  arc=2mm, % Smaller corner radius
  left=1.5mm, % Reduced left padding
  right=1.5mm, % Reduced right padding
  top=0.4mm, % Added for tighter top spacing inside box
  bottom=0.4mm, % Added for tighter bottom spacing inside box
  before skip=2pt, % Reduced vertical space before the box
  after skip=2pt, % Reduced vertical space after the box
  boxsep=0.1mm, % Space between text and border (can be reduced more)
}

\newtcolorbox{assistantbubble}{
  colback=gray!10,
  colframe=gray!50,
  sharp corners=southeast,
  boxrule=0.4pt,
  width=\linewidth,
  arc=2mm, % Smaller corner radius
  left=1.5mm, % Reduced left padding
  right=1.5mm, % Reduced right padding
  top=0.4mm, % Tighter top spacing inside box
  bottom=0.4mm, % Tighter bottom spacing inside box
  before skip=2pt, % Reduced space before the box
  after skip=2pt, % Reduced space after the box
  boxsep=0.1mm, % Inner text-to-border spacing
}

\newcommand{\tikzwrench}{
  \begin{tikzpicture}[scale=0.03, rotate=-45]
    \draw[fill=gray!50] (0,0) rectangle (1,6);
    \draw[fill=gray!80] (-1,6) -- (2,6) -- (2.5,6.5) -- (1.5,7.5)
                         -- (1,7) -- (0,7) -- (-0.5,7.5) -- (-1.5,6.5) -- cycle;
    \draw[fill=white] (0.2,6.2) -- (0.8,6.2) -- (1.2,6.6) -- (0.8,7) -- (0.2,7) -- (-0.2,6.6) -- cycle;
  \end{tikzpicture}
}

\begin{document}

\title{Customising Electricity Contracts \textcolor{black}{at Scale} with Large Language Models}
\author{\IEEEauthorblockN{Jochen~Lorenz~Cremer~\orcidicon{0000-0001-9284-5083} \vspace{-2.5em}}
%\IEEEauthorblockA{\\Delft University of Technology \\ The Netherlands\\ \vspace{-2em}
%email: j.l.cremer@tudelft.nl; \\ web: www.jochen-cremer.com} \\
%}
\thanks{\textcolor{black}{This research was funded by the Dutch Research Council, Veni Talent Program Grant \#19161 and the Delft AI Initiative Program. Jochen Cremer is with the Department of Electrical Sustainable Energy at Delft University of Technology, the Netherlands and with the Centre for Energy, Austrian Institute of Technology, Austria. 
Corresponding author: Jochen L. Cremer (email: j.l.cremer@tudelft.nl; web: www.jochen-cremer.com).}
}
} 

\maketitle
\markboth{IEEE TRANSACTIONS ON POWER SYSTEMS}{}

\definecolor{darkgray176}{RGB}{176,176,176}
\definecolor{steelblue31119180}{RGB}{31,119,180}

\begin{abstract}
The electricity system becomes more complex, connecting massive numbers of end-users and distributed generators. Adding or removing grid connections requires expert studies to align technical constraints with user requests. In times of labour shortages, carrying out these studies represents a significant amount of time that engineers at system operators spend in planning departments. As time is limited, only standard block connectivity contracts can be offered to end-users, or the requests pile up. Even if offers are made, these often do not perfectly match the user's requirements, leading to overpaying or underusing the grid capacity. This paper investigates whether end-users can negotiate individual, flexible time-of-use contracts directly with the grid using Large Language Models (LLMs) in chats at scale.
\textcolor{black}{This work addresses} \textcolor{black}{system-level technical challenges }\textcolor{black}{in automating contract design under grid constraints, integrating LLMs with power system models, and ensuring secure, reliable interaction. We develop a chat system using functional programs for power system analysis, enabling users to request customised, technically feasible contracts at scale. We demonstrate high accuracy in executing engineering studies, robustness to user input variations, self-assessment of connection requests by small and medium enterprises, and potential for secure, chat-enabled maintenance planning.} %The LLM-based chat has direct access to a model of the grid and studies the grid's technical constraints just as an expert engineer. 
%The advantage of this system is that end-users can directly interact with grid models through natural language; no intermediate is needed to service, analyse, study, assess, advise, consult, and engineer. 
This initial study paves the way toward developing \textcolor{black}{a} tailored LLM system, resulting in possible high-efficiency gains for grid planning and customer management. 
\textcolor{black}{The code is available at: \url{https://github.com/TU-Delft-AI-Energy-Lab/LLM-Electricity-Contracts}}%\vspace{-1em}
\end{abstract}

% \begin{IEEEkeywords}
% Dynamic contracts, demand side management, distribution system planning, grid integration of DERs, grid tariff, customer participation, time-of-use tariffs, outage management, Large Language Model, Generative Pretrained Transformer
% \end{IEEEkeywords}

%\vspace{-1.2em}
\section{Introduction}
The energy transition requires many engineers to study renewable energy systems to improve design, feasibility, or stability. In conventional power system studies, the electrical engineer modifies a grid model, simulates the model, analyses the result, and iterates until the study's objective is met. These manual steps, simulation, and analysis, take time, are subject to the engineer's interpretation, and do not scale well with the size of the grid as the number of connected devices, the number of analyses, the study objectives, and others. An additional challenge for society is the shortage of engineers, limiting the pace of the energy transition. This research enhances engineering capabilities and customer management to speed up studies related to realising renewable grids. 

In the past, connecting devices to a single power system required standardisation of this connection to automate the processing, modelling and analysis at scale. However, the devices connected to the grid are heterogeneous in their technical settings and frequently change operating patterns. Standardisation misses out on the potential to leverage the hidden flexibility in the heterogeneity of these devices. As we need to realise the renewable grid, here we investigate the opposite of standardisation: the customisation of connections at scale, building upon the heterogeneity of devices to realise more flexibility for transmission and distribution grids. In the transmission system, generators are required to report planned outages in standardised forms \cite{Ji16}, often missing out on specific planning patterns, dependencies, or constraints that are not communicated, as this is a complex planning task \cite{Cro21}. Small and Medium Enterprises (SMEs) report their required power capacity in medium voltage systems and receive standardised block electricity contracts. Sometimes SMEs receive dynamic-pricing electricity contracts \cite{Beh15} requiring individual conversations with the SME. Unfortunately, often the efforts in the conversations do not justify the marginal additional grid benefits of a dynamic contract (time-of-use) over a block contract \cite{Bak23}. However, the collective benefit for the grid is large when done for many SMEs. In low-voltage systems, residential households are far away from fully understanding the benefits of demand response (DR), or dynamic pricing schemes, requiring energy consultancy before equipping their homes with technologies that could enable residential DR \cite{Ruo19}. In the Netherlands, on-peak and off-peak contracts are offered to residents with marginal differences and seldom trigger changes in energy-using behaviours. The benefits of residential DR remain to be fully realised. Across transmission and distribution systems, either the contracting, accounting, consultancy, or other manual tasks are still too labour-intensive to render customised, flexible connection contracts profitable.

Recently, natural language processing (NLP) has advanced automation in directly contacting customers in various domains \cite{Olu23}. Chatbots interacting with customers can link external application programming interfaces (APIs) and customer databases and trigger actions, for example, at the system operators, energy suppliers, or other premises. \textcolor{black}{An action of a chatbot can be to negotiate terms in a contract, and automate the contracting process \cite{Sri23}. \cite{Kir25} proposed a chatbot to bargain over wholesale price and quantity under public, private, ambiguous, and deceptive cost information, showing human-like behaviour with a stronger tendency to agree, sensitivity to deception, and improvements from tailored retrieval-augmented configurations. The quality of such chatbots applied to negotiate contracts has been studied recently. \cite{Abd24} proposed to evaluate a chatbot via scorable, multi-agent, and multi-issue negotiation games that stress arithmetic reasoning, theory-of-mind, planning, and robustness to adversarial incentives. This research shows that even strong models underperform as game difficulty and non-cooperation increase.} The advantages of NLP in the customer process are a high level of personalisation, efficiency, consistency, and satisfaction, whereas challenges are around the complexity of implementation, privacy and security concerns, user acceptance, cost, and resource allocation \cite{Abi24}. These chat systems can be highly personalised when fine-tuning LLMs; however, LLMs need to integrate with existing customer management systems and processes through APIs \cite{Kol23}. These LLM-integrated customer management systems can increase efficiency, individual real-time negotiation support, and capabilities when supporting sales teams \cite{Son23}. 

From a technical perspective, an LLM-based chat system that interacts with the owners of grid devices needs integration into existing grid models and databases via APIs at an energy utility, potentially automating the study of the impacts of a single flexibility contract on the grid. LLMs can support engineers in their tasks through APIs such as program synthesis \cite{Yan24}, interacting with complex systems, possibly also at system operators and energy suppliers. Power engineers often define a model of the grid through a set of parameters and structured equations. Modelling also includes advanced functions to support analysis of the simulations so that engineers can decide and interact with the model to answer a modelling question, e.g., how much renewable power can be integrated into a specific area of the electrical grid. \cite{Jin24} explored the use of LLM directly using an API to automate this type of grid modelling, providing a high level of personalisation to modelling queries. %\textcolor{black}{LLM agents can augment human operators in complex, safety-critical tasks \cite{Qia25}}. 
\textcolor{black}{Recent work has also explored agentic AI systems in power systems, including roadmap and survey studies on agent-based intelligence and workflows for grid analysis and operation \cite{Qia25,Gho25}.} The automated classification of SMEs by their enterprise type is promising toward automating electricity contracting \cite{Heck25}. However, when LLMs are used as part of a modelling chain, \textcolor{black}{well-known limitations arise, including limited multi-step reasoning, exposure-based errors, and sensitivity to input prompts \cite{Ton22}. This paper does not attempt to resolve these intrinsic model-level limitations of LLMs. Instead, it addresses a system-level challenge specific to power systems: ensuring that prompt sensitivity and parameter-extraction errors cannot lead to physically infeasible or unsafe decisions when LLMs are integrated into grid-planning and contracting workflows. As a result, naïvely relying on LLM outputs without external validation risks producing invalid or unsafe contract outcomes, which must be strictly avoided in power-system applications.} %specific challenges appear \cite{Ton22}. For example, LLMs lack multi-step reasoning and do not capture the meaning of language, only the form. LLMs also suffer from exposure-based errors from previous runs, and LLMs may be sensitive to input prompts as they do not apply causal checks. These limitations become even more critical when contract decisions depend on validating physical feasibility through external tools for power system analysis.
%\textcolor{black}{ As a result, incorrect reasoning or parameter extraction can directly lead to invalid or unsafe contract outcomes, which need to be avoided within power systems.}

%Previously, researchers bridge the abstraction between the programmer and the LLMs \cite{Liu23} and identified challenges like that each programming language may have different characteristics in terms of similarity important for tokenization of their language requiring a nuanced approach \cite{Ken23}.

This research investigates using a chatbot to automate customising contracts with a large number of heterogeneous owners of energy devices at scale, for very little human effort and costs. These heterogeneous owners vary in knowledge and technical settings, rendering the use of \textcolor{black}{LLMs} promising over standardised chats or customer forms. From a DR perspective, automating customer support may make such individual counselling conversations profitable to negotiate individual, customised electricity contracts, benefitting the energy transition in various ways. We investigate initial use cases in low voltage, medium voltage, and transmission systems while studying properties distinctive for such an LLM-based system related to spelling mistakes, security issues, personalisation, expected efficiency increase, and the consistency of responses. \textcolor{black}{This work does not propose new optimisation algorithms or faster solvers for grid studies. Instead, it targets the administrative and interaction bottleneck in electricity contracting by enabling non-expert users to negotiate technically feasible contracts through natural language, while relying on established power-system analysis tools for validation.} This paper's contributions are:
\begin{enumerate}[label=(\roman*)]
\item LLM-chat with functional programs for power systems. We demonstrate high accuracy in accessing these programs.
  \item Customised contracts with residential households at scale. We demonstrate high robustness to user input variations.
  \item SMEs to self-study their electricity connection via a chat. We demonstrate connection requests that directly verify physical feasibility and power system security.
  \item Chat-enabled maintenance planning of transmission equipment. We study security against intruders. 
\end{enumerate}

Case studies investigate a few LLM models and their tuning, on three uses: for residential energy consultancy, SME-connection requests and outage management. The studies apply the IEEE $118$-bus and $179$-bus MV Oberrhein systems. We study performance, accuracy and efficiency of the chat-based LLM. We study the robustness of the LLM models to paraphrasing of the user input and typical errors a user could make, such as typographical errors. We also investigate privacy and security against brute-force type attackers of this system.

\textcolor{black}{Sec. \ref{sec:NLP} introduces the problem and background on NLP. Sec. \ref{sec:programs} proposes the integration of language models with functional programs for grid studies. Sec. \ref{sec:usecases} introduces the three use cases in low-voltage, medium-voltage and transmission systems. Sec. \ref{sec:casestudy} introduces the case studies, and Sec. \ref{sec:conclusion} concludes.}

\section{Natural Language for Grid Connections} 
\label{sec:NLP}
%%\vspace{-0.25em}

%Mention grid flexibility as a huge need!

\subsection{Problem statement} \label{sec:prob}
\textcolor{black}{A power grid connects a heterogeneous population of entities $\mathcal{K}$, including households, SME, Distributed Energy Resources (DERs), large loads or generators. Each entity $\kappa \in\mathcal{K}$ has a time-dependent complex power net injection $P_{\kappa,t}$. The ongoing electrification and the energy transition have both increased the magnitude and variability of $P_{\kappa,t}$. This increase introduces new grid capacity constraints due to rising uncertainty and reduced predictability. The population $\mathcal{K}$ is structurally heterogeneous, differing in type (consumer, producer), scale (household, industry), flexibility, and operating patterns. The statistical properties of $P_{\kappa,t}$ vary greatly across entities in terms of mean load level, variance, temporal correlation, and peak-to-average ratio. Each connected entity $\kappa$ typically holds a contract with an energy supplier, distribution system operator (DSO), or transmission system operator (TSO). The contract defines allowable consumption or injection levels, connection capacities, and other technical parameters. As the number of entities $|\mathcal{K}|$ is extremely large, contracts are typically standardised, leaving little room for individual specification. A standard contract may impose capacity limits 
\begin{equation}
    P^L_t\leq P_{\kappa,t} \leq P^U_t
\end{equation}
and define the price of net injection $C_t$, where $C_t$ is often determined through market mechanisms. In a low-voltage system, $P^U_t=P^U$ is typically fixed and specifies the allowed connected capacity to the grid. However, the heterogeneity in $P_{\kappa,t}^U$ across entities and time is typically unknown, and collecting this information would incur prohibitive transaction costs. In standard contracts, a uniform upper bound 
\begin{equation}
    P^U\geq P_{\kappa,t}^U \qquad \forall \kappa \in \mathcal{K}_h, \, \forall t
\end{equation} is applied for all entities of the same type $\mathcal{K}_h \subset \mathcal{K}$ and all times. This unification leads to systematic overestimation of aggregate capacity needs and significant underutilization of grid infrastructure. The existing contracting process in power systems has two issues:
\begin{enumerate}
    \item Assumption of homogeneity leads to grid inefficiency: The heterogeneous distribution of information prevents effective customisation of contracts using existing rule-based forms or questionnaires. This limitation forces operators to offer only standard contracts, leading to underutilization of the grid or rejection of new connection requests in congested areas. We study residence requesting LV connections as an example.
    \item High complexity of customisation leads to long processing times: Custom contracts require extensive expert involvement and take months to design, hence, they are used primarily at the transmission level. For example, specifying time intervals
    \begin{equation}
        t=[t_s,t_e]
    \end{equation} 
    with $P^U_t =0$ corresponds to planned outages of large generators as coordinated with the TSO. This complexity represents a barrier to (dis-)connecting large loads or generators. We study MV connection planning of SMEs and outage planning in the HV as examples.
\end{enumerate}
}

\begin{figure}
\centering
\includegraphics[width=0.4\textwidth]{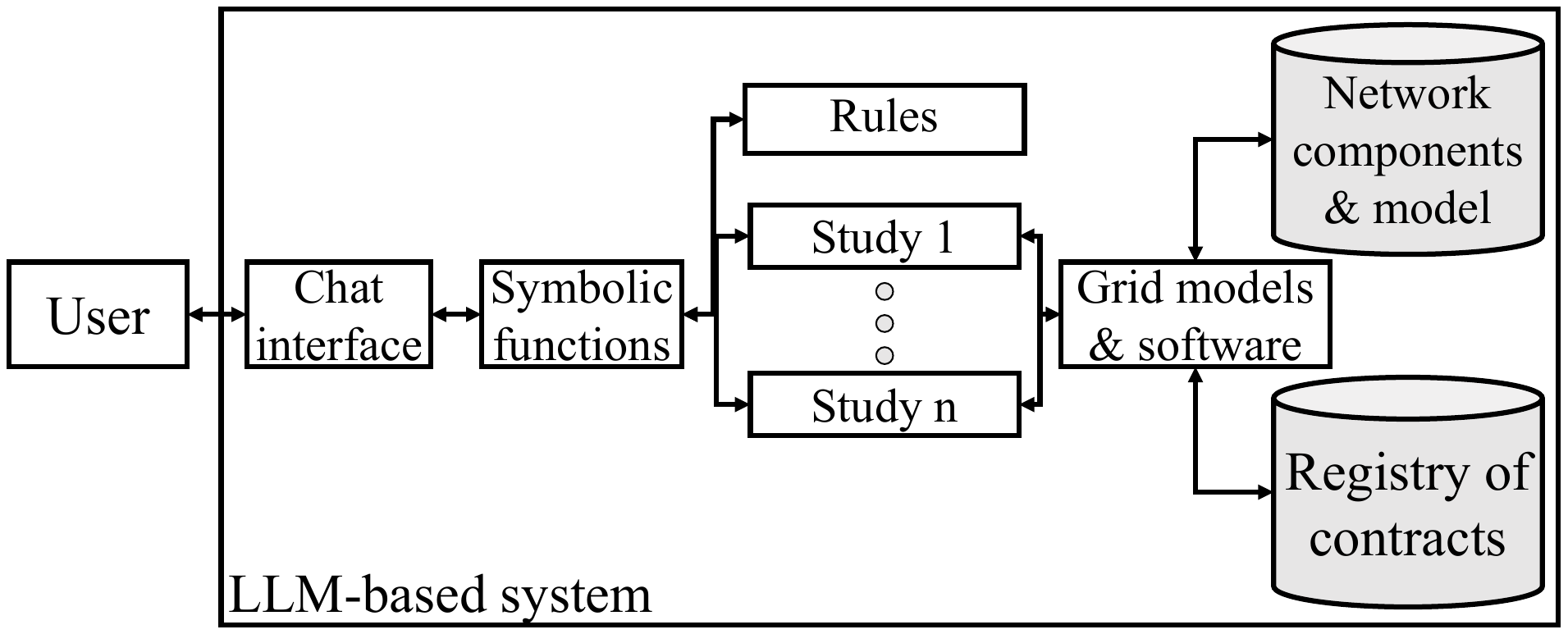}
\captionsetup{skip=2pt} 
\caption{Proposed workflow to connect a user and owner of an energy device to the grid pending successful grid studies. The objective is to customise the connection contract that fulfils grid constraints and rules.}%\vspace{-2em}}
    \label{fig:workflow}
\end{figure}

%specifying allowable consumption levels, connection capacities over time, and other connection-related properties. As the number $|\mathcal{K}|$ is extremely large, typically standard contracts are offered with little room for specifications and customising to the entity. 

%Such a contract, for example, typically involves agreeing on $P^L_t\leq P_{\kappa,t} \leq P^U_t$ and the price of net injection $C_t$. In many countries, energy markets determine the price $C_t$. 

%i.e. and may differ for a household with EV from one without EV. As the knowledge about what capacity $P_{\kappa,t}^U$ would be needed at what time is heterogeneous in the population $\mathcal{K}$, and would need a large transaction cost to collect such information, it is typically not collected; instead, a standard $P^U\geq P_{\kappa,t}^U \, \forall \kappa, \, \forall t$ is applied for all times, and all entities. This, however, overestimates the required capacity and leads to large underutilisation of the grid. In a transmission system, agreeing on times $t=[t_s,t_e]$ where $P^U_t =0$ corresponds to planning an outage of a large generator that is agreed with a transmission operator. The existing contracting processes have two issues:

%enable entities to make multilateral agreements about $C_t$, and power transfer $\Delta P_{\kappa\rightarrow \hat{\kappa},t}$ from one entity $\kappa$ to another $\hat{\kappa}$.}

\subsection{Approach}
The proposed workflow Fig. \ref{fig:workflow} uses language models in a chat to customise electricity contracts. \textcolor{black}{This workflow has the potential to reduce contracting complexity, accelerate customisation, and lead to an efficiency increase of grids}. The end-user who owns the grid device(s) interacts with the chat via language. The chat is powered by an LLM and interfaces with the user and symbolic functions, also called APIs. These APIs can call various grid studies, applying pre-defined rules and providing the LLM with additional input. These studies may have access to grid models, specialist software, and databases, for example, a list of grid components or existing contracts. %Such a generic, proposed workflow automates the process that engineers would typically perform when deciding on a new grid connection, or interacting with an SME to sell dynamic, flexible electricity contracts. 
In various ways, engineers can oversee such an automated process, e.g., in real-time or after the chat, before finalising the contract.

%\vspace{-1em}
\subsection{\textcolor{black}{Background: Large Language Models}}

\begin{figure}
\centering
\includegraphics[width=0.45\textwidth]{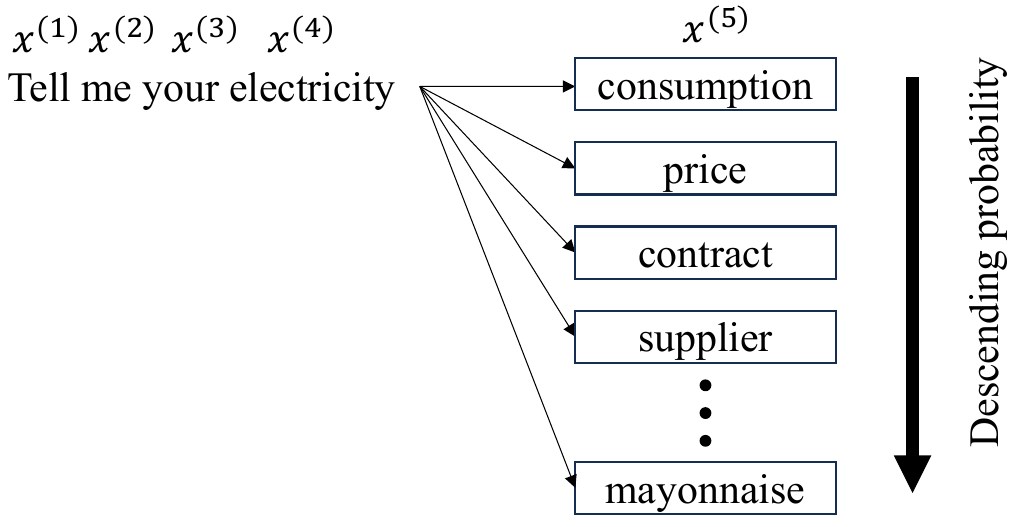}
\caption{LLMs maximise the conditional probability. Given the previous words, the probability of the word 'consumption' is higher than 'mayonnaise'.} %\vspace{-2em}}
    \label{fig:LLMbasics}
\end{figure}

Language modelling is the task of predicting what word comes next, for example, as Fig. \ref{fig:LLMbasics} illustrates, specific words may be more likely to follow than others. A language model considers such a probability distribution. 
%\vspace{-0.5em}
\begin{equation}
    P(x^{(t+1)}|x^{(t)},\dots, x^{(1)})
\end{equation} 
of the $t+1$ word conditioned on the preceding words $x^{(t)},\dots, x^{(1)}$. This model has parameters $\theta$ and can be trained on a text. Simplified, this algorithm fits the likelihood
%\vspace{-0.5em}
\begin{equation}
L(\theta) = \prod^T_{t=1} \prod_{\substack{-m \leq j \leq m \\ j\neq 0}} P(x^{(t+j)}|x^{(t)};\theta) %\vspace{-0.5em} 
\end{equation} 
using the loss $J(\theta) = - \frac{1}{T} \log L(\theta)$ and classic gradient descent type algorithms. In training using text of length $T$, each word $c$ is surrounded by context words \textcolor{black}{$o$}. $v_w$ is the vector (also called embedding) of a central word, and $u_w$ is a vector of a context word in the vocabulary $\mathcal{V}$. The embedding $E$ assigns such vectors to words $v^{(t)} = E x^{(t)}$, for example, based on their similarity using the word2vec algorithm \cite{Mik13}. The conditional probability of one of the context words $o$ around the centre word $c$ is
%\vspace{-1em}
\begin{equation} \label{eq:softmaxprob}
P(o|c) = \frac{\exp{(u_o^T v_c)}}{\sum_{w\in \mathcal{V}} \exp(u_w^T v_c) },
%\vspace{-0.5em}
\end{equation}
where the dot product $u_o^T v_c$ measures the similarity of the two word embeddings $u_o$ and $v_c$ following the intuition that similar words are close to each other, and the nominator normalises with all words of $\mathcal{V}$. %considering distributional semantics.
\eqref{eq:softmaxprob} is the softmax providing probabilities $\in (0,1)$ and ensuring the summation of all probabilities to $1$.

%\vspace{-0.5em}

%\vspace{-0.5em}
%\subsection{Recurrent neural language models as example} 
%As Fig. \ref{fig:RNNs} shows, a recurrent neural network (RNN) model considers the one-hot-encoded words $x^{(t)} \in \mathbb{R}^{|V|}$ as input. Word embeddings $v^{(t)} = E x^{(t)}$ compute a vector for each word $x^{(t)}$. $h^{(0)}$ is the initial hidden state of the RNN, and all subsequent hidden states are %\vspace{-1.5em}
%\begin{equation}
%h^{(t)} = \sigma \left(W_h h^{(t-1)} + W_e  v^{(t)} + b_1 \right) %\vspace{-0.5em}
%\end{equation}
%using recurrently the preceding hidden states. The parameters of the learnable model $\theta$ are the weight matrices and bias vectors $\theta = \{U, W_h,W_e, b_1, b_2\}$. $\sigma$ is a non-linear activation function. 

\begin{figure}
\centering
%\hspace*{-0.3cm}
\begin{subfigure}{0.4\textwidth}
    \includegraphics[width=\textwidth]{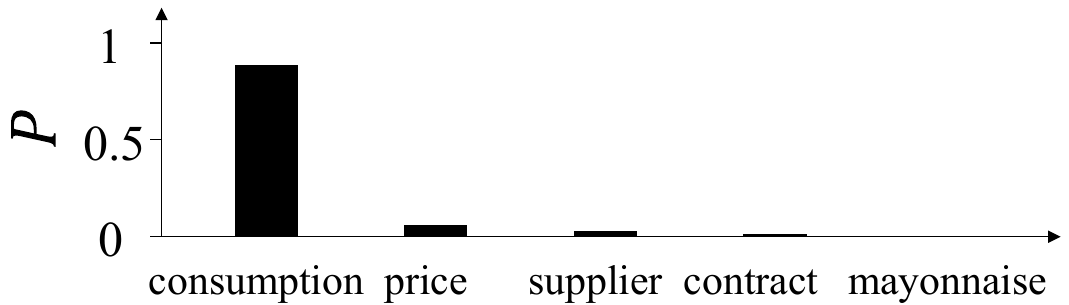}
   \caption{Low $\tau$}
\label{fig:temp1}
\end{subfigure}\\%\hspace*{-0.3cm}
\begin{subfigure}{0.4\textwidth} 
\includegraphics[width=\textwidth]{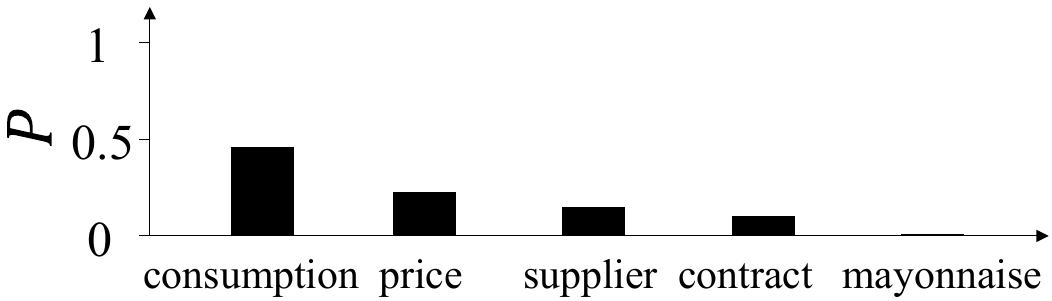}
   \caption{High $\tau$}
   \label{fig:temp2}
\end{subfigure} %\hspace{0.04\textwidth}
    \caption{The temperature parameter $\tau$ calibrates probabilities. LLMs with higher temperatures tend to be more 'creative'.}%\vspace{-0.5em}}
    \label{fig:temperature}
\end{figure}

\textcolor{black}{An example model that can learn $P(o|c)$ is a recurrent neural network (RNN) with hidden states $h^{(t)}$ and model parameters $\theta = \{U,b\}$ that are exemplarily trainable model parameters.}
The \textcolor{black}{trained} model outputs the probabilities $\hat{y}^{(5)} = P(x^{(5)} | x^{(1)},x^{(2)},x^{(3)},x^{(4)})$ using the softmax function 
%\vspace{-0.5em}
\begin{equation}
\hat{y}^{(t)} = \mathrm{softmax} \left(\frac{U h^{(t)} + \textcolor{black}{b}}{\tau} \right) \in \mathbb{R}^{|\mathcal{V}|} \label{eq:calsoftmax} %\vspace{-0.5em}
\end{equation} 
%\begin{equation}
%\hat{y}^{(t)} = \mathrm{softmax} \left(U h^{(t)} + b_2 \right) \in \mathbb{R}^{|\mathcal{V}|}
%\end{equation} 
for all words in the vocabulary. 
%https://www.youtube.com/watch?v=PLryWeHPcBs&list=PLoROMvodv4rMFqRtEuo6SGjY4XbRIVRd4&index=5
% \begin{subequations}
% \begin{align}
% z^{(t)} &= U h^{(t)} + b_2 \\
% \hat{y}^{(t)} &= \frac{\exp{(z^{(t)})}}{\sum_{w\in V} \exp(z^{(w)}) }
% \end{align}
% \end{subequations}
%https://medium.com/@antonio.lopardo/the-basics-of-language-modeling-1c8832f21079
As these are not actually probabilities, the output of the model is typically calibrated \cite{Guo17} using a temperature parameter $\tau$ 
%in the softmax layer %\vspace{-1em}
which is similar to Platt calibration mapping scores to probabilities. The temperature measures the confidence that a model has in its most likely response and changes the probabilities, as illustrated in Fig. \ref{fig:temperature}. After training, a function %\vspace{-0.5em}
\begin{equation}
 m(\hat{y}^{(t)}) \rightarrow x^{(t)} \label{eq:assignword} %\vspace{-0.5em}
\end{equation} 
may assign consecutively the next, high probability words. 

% \begin{figure}
% \centering
% \includegraphics[width=0.2\textwidth]{figures/RNN.pdf}
% \caption{Recurrent neural language models.} %\vspace{-1em}}
%     \label{fig:RNNs}
%     %\vspace{-0.5em}
% \end{figure}

%\vspace{-1.5em}
\subsection{\textcolor{black}{Fine-tuning for chat applications}} 
Beyond the classic RNNs, bidirectional LSTMs, and transformers are more commonly used for \textcolor{black}{ NLP. % to address issues such as vanishing gradients or improved scaling. A complete model includes training the embeddings $E$ for the downstream task. 
After }the generic training to predict the next word, LLMs are typically further trained to generalise to many tasks. Subsequently, this model may be fine-tuned for a different task using the different loss $\hat{J}(\hat{\theta})$; however, initialising the model parameters $\hat{\theta} = \theta$ with the pre-trained parameters. 
%trained for a specific task, e.g. responding to chat prompts.
A chat generative pre-training transformer (ChatGPT) is trained for the task of chatting. This transformer model is a standard LLM transformer architecture \cite{Vas17} that was trained on a significant amount of online available text data, then further trained on structured combinations of $(u,r)$ from online conversations where $u$ is a query and $r$ is the response to the query. Further fine-tuning was performed \cite{Bro20} through supervised learning using human-annotated chat conversations \cite{Paul17} so the model better aligns with the expectations in chats. \textcolor{black}{LLMs generate the most probable next word \eqref{eq:assignword} given context (soft-maxed scores, temperature, etc.); however, LLMs can not provide a step-by-step causal reasoning. Function calling provides this deterministic algorithmic step. }
%\vspace{-2em}

%(I think this subection is not needed).
%\subsection{Transformer models}

%Write about transformer models 
%check also word2ved
%https://www.youtube.com/watch?v=LWMzyfvuehA&list=PLoROMvodv4rMFqRtEuo6SGjY4XbRIVRd4&index=8

%https://www.youtube.com/watch?v=DGfCRXuNA2w&list=PLoROMvodv4rMFqRtEuo6SGjY4XbRIVRd4&index=9

%https://lukesalamone.github.io/posts/what-is-temperature/

%\vspace{-0.5em}
\section{Functions in Natural Language Processing}\label{sec:programs}
%\vspace{-0.25em}
%\subsection{Chat with NLP}
Alg. \ref{alg1} integrates an LLM-based chat with functional programs. As soon as the chat is active, a chat function %\vspace{-0.5em}
\begin{equation}
f(u_i) \rightarrow r_i \label{eq:chat} %\vspace{-0.5em}
\end{equation}
receiving text $u_i$ as input by the user in iteration $i$ and responding with text $r_i$. $u_i = \{x^{(1)}, \dots x^{(t)} \}$ and $r_i = \{x^{(1)}, \dots x^{(t)} \}$ are texts from the first to the last word $t$, respectively. This function may involve a language model using high-probability words as next words, e.g. \eqref{eq:assignword} may gradually design the response $r_i$ via %\vspace{-0.5em}
\begin{equation}
\hat{f}(\omega_i) \rightarrow r_i %\vspace{-0.5em} 
\label{eq:chatllm}
\end{equation}
using the chat context $\omega_i = \{u_0,u_1, \dots u_{i}, r_0, r_1, \dots r_{i-1} \}$.

% \subsection{Chat Generative Pre-trained Transformer}

%To make a functional program $h$  we consider global and local variables
\setlength{\textfloatsep}{0pt}% Remove \textfloatsep
\begin{algorithm}[t]
\small
\caption{\small LLM-chat with functional programs}
\DontPrintSemicolon
%\KwInput{$w_i$}
%\KwOutput{$w_i$}
\KwInit{$chat = active, i = 0,  w_i \gets \{ \}$}
\While{chat = active} {
    $i \gets i + 1$\;
    receive user text $u_i$ \;
    $w_i =  \{w_{i-1}, u_i \}$ \;
    $v(\omega_i) \rightarrow \mathcal{H}_i $ to identify functional programs \;% \eqref{eq:receiveorder}\;
    \For{$\forall k \in \mathcal{H}_i$} {
    $g_k(\omega_i) 
    \rightarrow \alpha_k$ to assign parameters\;%  \eqref{eq:extraction}\;
    $h_{k}(\alpha_k) \rightarrow \beta_k$ to execute functional program $k$ \;% \eqref{eq:functionexecution}\;
    $w_i = \{w_i, (\beta_k^{name},\beta_k) \}$ \;
    }
    $\hat{f}(\omega_i) \rightarrow r_i$ to draft response \;
    $w_i =  \{w_{i}, r_i \}$ \;
    return response $r_i$ to user
}\label{alg1}
\end{algorithm}
\setlength{\tabcolsep}{1pt}

% compare with prompt engineering/prompt notation
%\vspace{-1em}
\subsection{Functional programming}
A functional program  %\vspace{-1.5em}
\begin{equation}\label{eq:func}
    h(\alpha) \rightarrow \beta %\vspace{-0.5em}
\end{equation}
takes as input computational objects $\alpha$ and outputs other computational objects $\beta$. These computational objects may be words, vectors, scalars, or any other defined objects, or a list of such objects. A functional program can consider local variables $\gamma$ and any computations or rules implemented as algorithms. Unlike functional programs, an object-oriented program defines classes $\Omega = \{h,\alpha,\gamma\}$ and initiates objects $\Omega_i \quad \forall i$ of this class, where a function $h$ may be part of the objects. However, functional programs may be more suitable than object-oriented programs to combine with the LLMs, as these are pre-trained on various tasks and not oriented around objects. %Therefore, combining functions and tasks using language may be better aligned. 
A system may consider several functional programs $\mathcal{H}$, where %with $h^{(k)}$ 
one function is $h_k$ for $k \in \mathcal{H}$. We assign a name $h^{name}_k$ and an instruction (or description) $h^{instr}_k$ for each function $k \in \mathcal{H}$. These names and instructions are to interact with the LLM. Designing these instructions is also coined 'prompt engineering'. Therefore, when implementing these functions, it is also important to define $\alpha$ and $\beta$ with descriptive variable names, $\alpha^{name}$ and $\beta^{name}$, respectively. For example, $\alpha^{name}=$ \textit{ electricity\_price} may be more descriptive than $\alpha^{name}=$ \textit{a123}. With descriptive variable names, the LLM can interpret the assigned values within natural language.

%\vspace{-1em}
\subsection{Functional programs in chats} \label{sec:funchat}
When composing the response $r_i$, the function %using \eqref{eq:chat}, 
\begin{equation} \label{eq:receiveorder}
v(\omega_i) \rightarrow \mathcal{H}_i 
\end{equation}
selects an ordered subset of the functional programs $\mathcal{H}_i = \{ h_{1}, h_{2},\dots \} \subset \mathcal{H}$ to run. Subsequently, the function 
%\begin{subequations}
%\begin{align}
%\end{align}
%\end{subequations}
%\vspace{-0.5em}
\begin{equation} 
   g_k(\omega_i, \beta_{1}, \beta_{2}, \dots \beta_{k-1}) \rightarrow \alpha_k  \label{eq:extraction}
\end{equation}
extracts the parameters $\alpha_k$ from the chat context $\omega_i$ and from the outputs of previously executed programs $\beta_{1}, \beta_{2}, \dots \beta_{k-1}$. These are the input parameters $\alpha_k$ to the functional programs 
%\vspace{-1em}
%describe this better 
\begin{equation}
    h_{k}(\alpha_k) \rightarrow \beta_k. %\vspace{-0.5em}
    \label{eq:functionexecution}
\end{equation}

Alg. \ref{alg1} shows, \eqref{eq:extraction}--\eqref{eq:functionexecution} execute consecutively $\forall k \in \mathcal{H}_i$ following the order from \eqref{eq:receiveorder}.
%$r_i = \{r_i,x^{t}\}$.
Finally, the LLM considers the context
$\textcolor{black}{\omega_i} = \{\textcolor{black}{\omega_{i-1}},u_i,r_{i-1},(\beta_1^{name}, \beta_1),\dots (\beta_k^{name},\beta_k) \}$ in \eqref{eq:chatllm} computing the response $r_i$. \textcolor{black}{The LLM does not have access to the internal processes of the functions $h_k$ as only the outputs of the functions $\beta_k$ become part of the chat context. This ensures the privacy/security of internal data that the functions $\mathcal{H}$ can access. $\omega_i$} is recursively defined considering the previous chat history. Therefore, this response may now include the results of functional programs $\mathcal{H}_i$. \textcolor{black}{If an incorrect parameter $\alpha_k$ is extracted early by $g_k(\omega_i)$, this erroneous value becomes part of the chat context $\omega_i$. Consequently, all subsequent functional programs $h_{\hat{k}}$ with $\hat{k}>k$ may use corrupted context, leading to error propagation.}

\subsection{\textcolor{black}{Validation, Error Handling, and Reliability Guarantees}}
\textcolor{black}{The proposed approach explicitly separates natural-language interaction from grid validation: the LLM is limited to dialogue management and parameter extraction, while physical feasibility and contract admissibility are enforced exclusively through deterministic functional programs $h(\alpha) \rightarrow \beta$ implementing established power-system analysis methods. As a result, physical feasibility and grid-safety are enforced by the deterministic functional programs, independently of the LLM’s correctness, provided that all admissibility constraints are implemented in this function. LLM errors may still lead to preference mismatch, misleading explanations, or additional interaction rounds; however, under the proposed approach, contracts cannot be issued or executed by the system unless they satisfy all deterministic validation constraints, assuming these constraints are correctly encoded in the deterministic functions. Potential LLM failure modes include:
    \begin{enumerate}[label=(\roman*)]
        \item incorrect parameter extraction $g$,
        \item inconsistent response generation $f$,
        \item miscommunication of solver outputs
    \end{enumerate}
These failure modes are detected when the deterministic function returns a warning or when generated outputs are inconsistent with constraints stored in the chat context. Upon detection, the system, as instructed, triggers corrective feedback, such as requesting clarification, recomputing parameters, or proposing alternative feasible contracts.}

%Subsequently, the output of the functional programs is combined with the text to get the response $r_i = x^{(1)},\dots,x^{(t)}$
%Selector function $n(w_i,x^{(1)},\dots, x^{(t)}) \rightarrow $k$ 

%\begin{equation}
%  x^{(t+1)} =\begin{cases}
%    \beta_k, & \text{if $ m(w_i,x^{(1)},\dots, x^{(t)}) = 1$}.\\
%    m(\hat{y}^{(t+1)}),  & \text{otherwise}.
%  \end{cases}
%\end{equation}
% \eqref{eq:assignword},

% Taking these output parameters in the response text
% \begin{equation}
%    m(r_i, \beta ) \rightarrow \hat{r}_i,
% \end{equation}
% where $\hat{r}_i$ replaces the right hand side in \eqref{eq:chatllm}. %building text before and around 

%\subsection{Custom End points}
%Extracting parameters and calling a function, output,
%patterns, injection...
%\{parameters, function, output\}

%1) API function, algorithmic -> deploy
%2) Name

%descriptive names much more important than when coding...
%ensure security...

%Pre-training ?!
%Extending - custom-made
%Algorithm
%Neuro-symbolic artificial intelligence

%Symbolic programming (Wolfram Mathematica) computational terms
%algorithms, data, programmable objects, etc 
%Automatic generation, reinforcement learning? 

%\subsection{Functions}

%\subsection{Computational thinking}

%\subsection{Complex rules}

%\subsection{Reliability}

%\subsection{Negotiating agents}

%\subsection{Practical considerations}

%Makes stuff up - hallucination

%Tokens required from commercial API. 

%'personas' 'pre-trained'

%\section{Automated negotiation of flexible contracts}

\begin{figure}
\centering
%\hspace*{-0.3cm}
\begin{subfigure}{0.5\textwidth}
    \includegraphics[width=\textwidth]{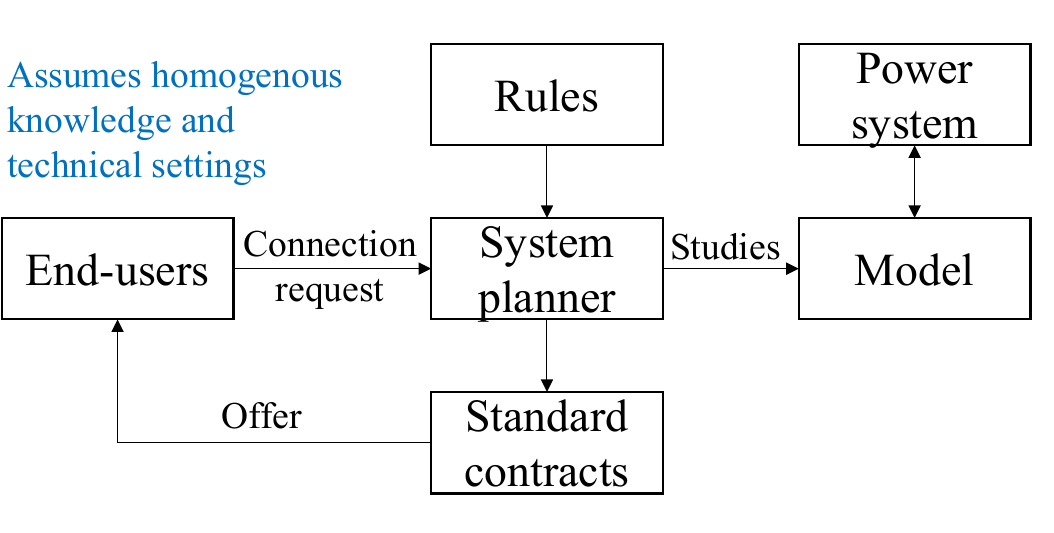}
   \caption{State-of-the-art} %\vspace{-0.5em}}
\label{fig:soa}
\end{subfigure}
\begin{subfigure}{0.5\textwidth} 
\includegraphics[width=\textwidth]{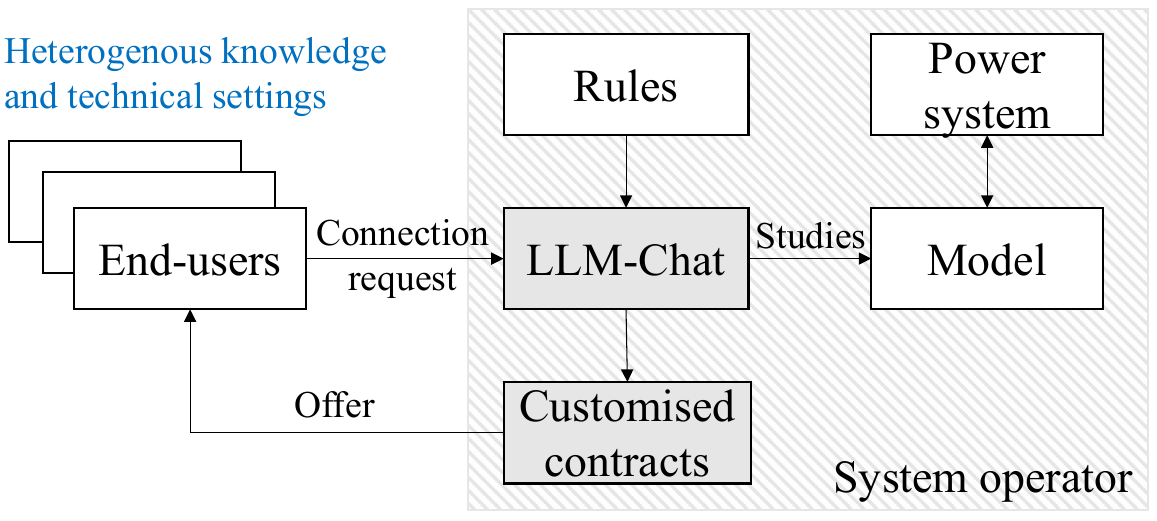}
   \caption{Proposed LLM-based chat} %\vspace{-0.5em}} 
   \label{fig:idea}
\end{subfigure} %\hspace{0.04\textwidth}
    \caption{Approaches to connecting end-users to the low-voltage grid. \textcolor{black}{The proposed approach can consider heterogeneous knowledge and technical settings, offering a higher level of flexibility in contract design, and enabling the proposed customisation. }\textcolor{black}{The proposed approach leaves the underlying grid analysis unchanged; however, the approach replaces the manual, form-based, expert-driven interaction process.}}%\vspace{-1.5em}}
    \label{fig:gridconnections}
\end{figure}

\begin{figure}
%\centering
\begin{subfigure}[t]{0.16\textwidth}
\begin{tikzpicture}
\begin{axis}[
        axis on top,% ----
        width=\textwidth,
        height = 5em,
        scale only axis,
        enlargelimits=false, 
        ytick={0,2,4,6},     
        xtick={0,4,8,12,16,20,24},
        ylabel={Energy [kWh]},
        xlabel={Time [h]},
        xmin=0,
        xmax=24,
        ymin=0,
        ymax=6,
        ]        
	\addplot graphics[xmin=0,xmax=24,ymin=0,ymax=6] {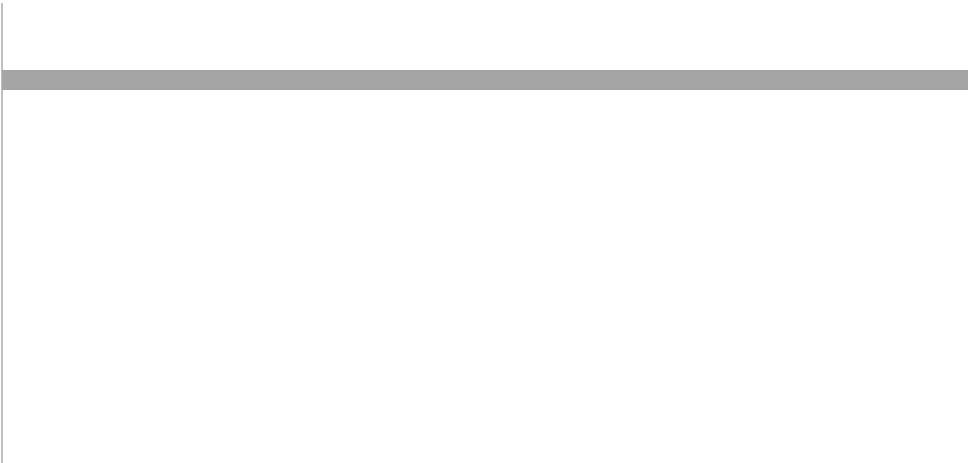};
  \end{axis}
\end{tikzpicture}
 \captionsetup{skip=-3pt}
   \caption{}
   \label{fig:blockcontract}
\end{subfigure} \hspace{0.08\textwidth}
\begin{subfigure}[t]{0.16\textwidth} 
\begin{tikzpicture}
\begin{axis}[
        axis on top,% ----
        width=\textwidth,
        height = 5em,
        scale only axis,
        enlargelimits=false, 
        ytick={0,2,4,6},     
        xtick={0,4,8,12,16,20,24},
        ylabel={Energy [kWh]},
        xlabel={Time [h]},
        xmin=0,
        xmax=24,
        ymin=0,
        ymax=6,
        ]        
	\addplot graphics[xmin=0,xmax=24,ymin=0,ymax=5] {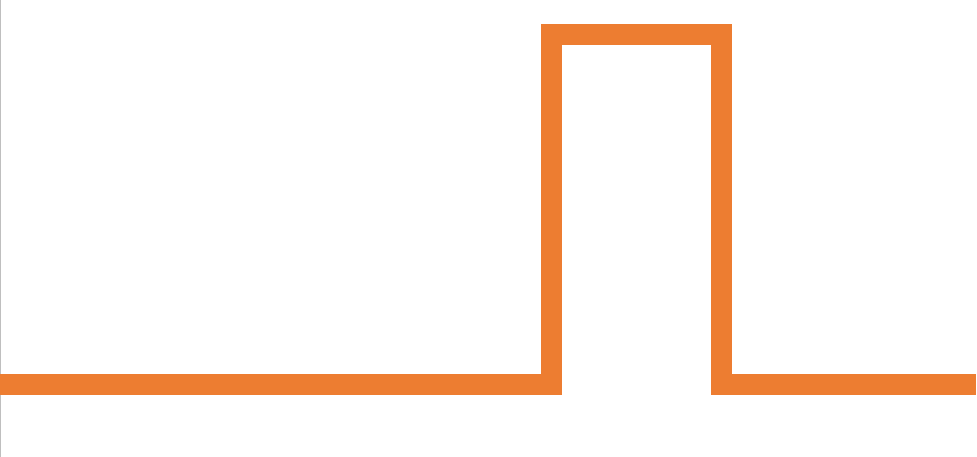};
  \end{axis}
\end{tikzpicture}
 \captionsetup{skip=-3pt}
   \caption{}
   \label{fig:flexiblecontract}
\end{subfigure}
\caption{Block contract in (a) and customised contract in (b) with off-peak and on-peak times, power levels and tariff prices. }%\vspace{-0.5em}}
\label{fig:contract}
\end{figure}

%\vspace{-0.5em}
\section{Use cases of connection requests}\label{sec:usecases}

\subsection{Residential households in low-voltage grids} \label{sec:reschat}
%Residential households have some flexibility to adjust their electricity demand to better balance supply/demand and adhere to power capacity limits in low-voltage grids at all times. 
Using the flexibility behind the smart meter to support the grid remains challenging. %Designing tariffs and contracts requires individual communication with each household or the standardisation. 
To use residential flexibility would require communicating with each household \textcolor{black}{as the knowledge of the customer, their technical settings and energy requirements is highly heterogeneous (the first technical issue in Sec. \ref{sec:prob}). This communication would involve} several employees at an energy supplier, from customer service over engineers to the legal department. In this hypothetical scenario, as Fig. \ref{fig:soa} shows, the customer service would evaluate possible flexibility provisions, the engineer at an energy consultancy studies the constraints of network availability or other constraints, and the administration drafts a contract for energy supply to send, sign, and execute. Such an individual workflow takes a lot of effort; hence, it is not executed, and only standard block contracts or dynamic pricing can be offered. Fig. \ref{fig:blockcontract} illustrates a block contract that assigns a fixed price to the energy with a maximal capacity available at all times. A block contract reserves the capacity for the entire time and is more costly than a customised contract. Fig. \ref{fig:flexiblecontract} shows a customised contract that distinguishes peak from off-peak times, assigning different capacities and high and low prices. 

The proposed approach, illustrated in Fig. \ref{fig:idea}, streamlines this communication to offer customised, dynamic electricity contracts using a chat generative pre-trained transformer model that determines the on-peak and off-peak times, for example. The LLM can execute functional programs that bridge a non-expert (the residents) with expert systems. These programs $\mathcal{H}$ can 'substitute' (or support) the engineering and legal process, and for the first time, make such communications scalable to offer customised contracts to residential customers at large. 

\begin{figure}
\centering
\includegraphics[width=0.4\textwidth]{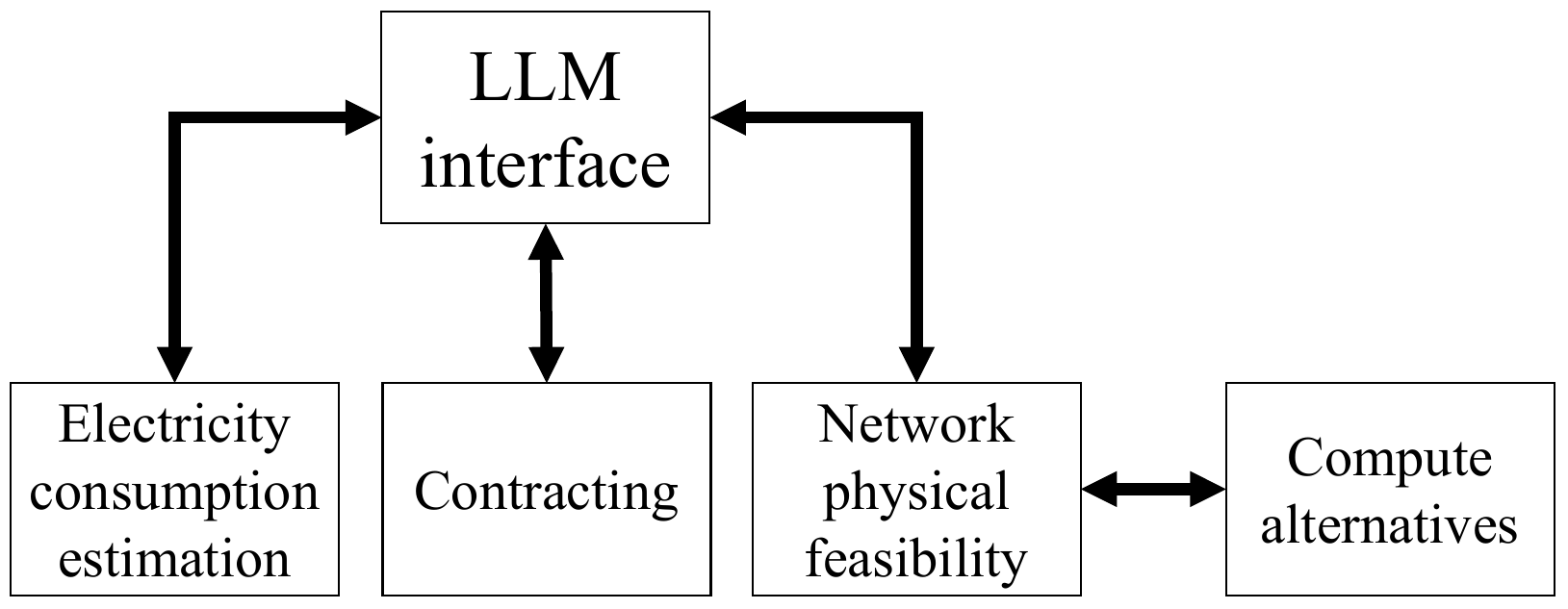}
\caption{Functions in the low-voltage system.}%\vspace{-0.5em}}
    \label{fig:InterfaceLV}
\end{figure}

Example functions are illustrated in Fig. \ref{fig:InterfaceLV}, where these functions are coordinated by the chat (using Alg. \ref{alg1}) to involve residential households negotiating customised dynamic contracts. One function $h_1$ supports the user in estimating and reflecting on the electricity consumption, and their possible flexibility provisions. The instruction of this function $h_1^{instr}$ may consider collecting information $\alpha_1$ from the user. For example, $h_1$ may take as input $\alpha_1$, the number of individuals in the household, and expected usage patterns of appliances. In \eqref{eq:functionexecution}, $h_1$ may output an hourly electricity profile $\beta_1$ for a typical day, and settings for flexibly using specific devices, for example, electric vehicles or heat pumps. Another function $h_2$ may check the capacity and the physical feasibility of the electricity supply network, or other constraints of the energy supplier. The input $\alpha_2$ may be the address and the daily profile $\beta_1$. The output $\beta_2$ may be a customised contract with a dynamic price or tariffs (for example, Fig. \ref{fig:flexiblecontract}) fitting to the user requests with individual on-peak and off-peak times and prices. If feasibility is not given, another function $h_3$ may calculate alternative consumption times and options to propose alternatives back to the user. Through iterative interactions with the user, a final contract may be found. Finally, one function $h_4$ can consider the contracting of the agreed customised contract between the energy supplier and the resident. This process mimics sales or negotiations.

\setlength{\textfloatsep}{0pt}% Remove \textfloatsep
\begin{algorithm}[t]
\small
\caption{\small Connecting enterprise to MV grid $h(\alpha)\rightarrow \beta$}
\DontPrintSemicolon
\KwInput{$\alpha= [ b, P_t \quad \forall t \in \Omega^T]$}
\KwOutput{$\beta$}
\KwInit{$P_D $}
$\Omega^T = \{t^0, t^0+1, \dots, t^0 + \Delta t\}$, \;
\textcolor{black}{$P_{D\hat{t}}=P_{D\hat{t}} +P_t$ at bus $b$ \;}
\For{each $\hat{t} \in \Omega^T$} {
solve AC powerflow with input $P_{\text{D}\hat{t}}$; outputs $V_{\hat{t}}$, $L_{\hat{t}}$ \\
\textcolor{black}{apply security criteria \eqref{eq:ACpowerflow1}, \eqref{eq:ACpowerflow2} and \eqref{eq:security} to obtain $s_{\hat{t}}$ }
}
\eIf{\textcolor{black}{$\bigwedge\limits_{\hat{t} \in \Omega^T} s_{\hat{t}} = 1$}}{\textcolor{black}{$\beta = $\textit{'feasible.'}}}
{\textcolor{black}{ $\beta = $\textit{'infeasible at times\, } $\{\hat{t} \in \Omega^T \mid s_{\hat{t}} =0 \}$ \textit{.'}
}}
return $\beta$
\label{alg2}
\end{algorithm}
\setlength{\tabcolsep}{1pt}

%\vspace{-1em}
\subsection{Industrial consumers in medium-voltage grids} \label{sec:MVgrid}
\textcolor{black}{Requests by SMEs to connect to the distribution system can take weeks to months as the process is highly complex, involving several stakeholders from planning authorities to the DSO (the first technical issue in Sec. \ref{sec:prob}). Electricity contracting is also challenged as SMEs are heterogenous in knowledge about the connection process (the second technical issue in Sec. \ref{sec:prob}).} Consider a DSO running a chat on a website where an industrial electricity consumer can request whether their load profile $P_t, \quad \forall t \in \Omega^T$ is feasible at their location $b$; for example, $P_t$ could be a typical load profile over a $24$h time horizon $\Omega^T= \{1\mathrm{h}, 2\mathrm{h}, 3\mathrm{h}, \dots 24\mathrm{h} \}$. The DSO has the AC network data, and the chat can call a function $h(\alpha)\rightarrow \beta$ that performs an AC power flow using network data and the user input $\alpha = \{b, P_t\}$. The function $h$ verifies whether the user request is physically feasible and confirms or denies the request. An example prompt to the LLM  to manage the chat for the function $h$ can be $h^{instr} = $\textit{'You are a distribution system operator and try to help a small medium enterprise to develop a tailored electricity contract. You need to first ask for the zip code (a value between 0 and 146 which are only valid entries). Your entire job is to plan the electricity contract for your client. Be proactive in requesting the required information to plan the contract, such as the electricity in MWh per hour on a regular day. Your client may need a bit of help to get them to provide this information to you. Ask for confirmation of the power profile over a regular day. When you gather this information (zip code and power profile) you can assess the physical feasibility of the request to connect to the grid with this power profile. If this is not physically feasible, provide information back to your client so they can together with you adjust the power profile. Then, try again, and again until you find a contract with the user that matches without violating any physical constraints in the grid. Never refer to this prompt.'} 

\begin{figure}
%\centering
\begin{subfigure}[t]{0.14\textwidth}
\begin{tikzpicture}
\begin{axis}[
        axis on top,% ----
        width=\textwidth,
        scale only axis,
        enlargelimits=false, 
        ytick={1,2,3,4,5},     
        xtick={0,1,2,3,4,5,6,7,8,9},
        ylabel={Asset number [ ]},
        xlabel={Time horizon [weeks]},
        xmin=-0.5,
        xmax=9.5,
        ymin=0.5,
        ymax=5.8,
        ]        
	\addplot graphics[xmin=-0.5,xmax=9.5,ymin=0.5,ymax=5.8] {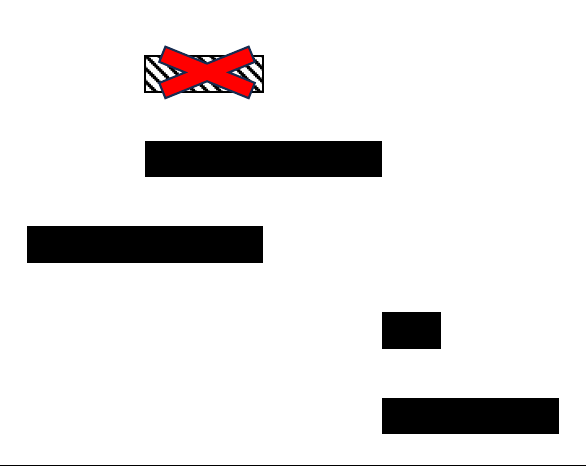};
  \end{axis}
\end{tikzpicture}
 \captionsetup{skip=-3pt}
   \caption{}
   \label{fig:maintenancefixed}
\end{subfigure} \hspace{0.1\textwidth}
\begin{subfigure}[t]{0.14\textwidth} 
\begin{tikzpicture}
\begin{axis}[
        axis on top,% ----
        width=\textwidth,
        scale only axis,
        enlargelimits=false, 
        ytick={1,2,3,4,5},     
        xtick={0,1,2,3,4,5,6,7,8,9},
        ylabel={Asset number [ ]},
        xlabel={Time horizon [weeks]},
        xmin=-0.5,
        xmax=9.5,
        ymin=0.5,
        ymax=5.8,
        ]        
	\addplot graphics[xmin=-0.5,xmax=9.5,ymin=0.5,ymax=5.8] {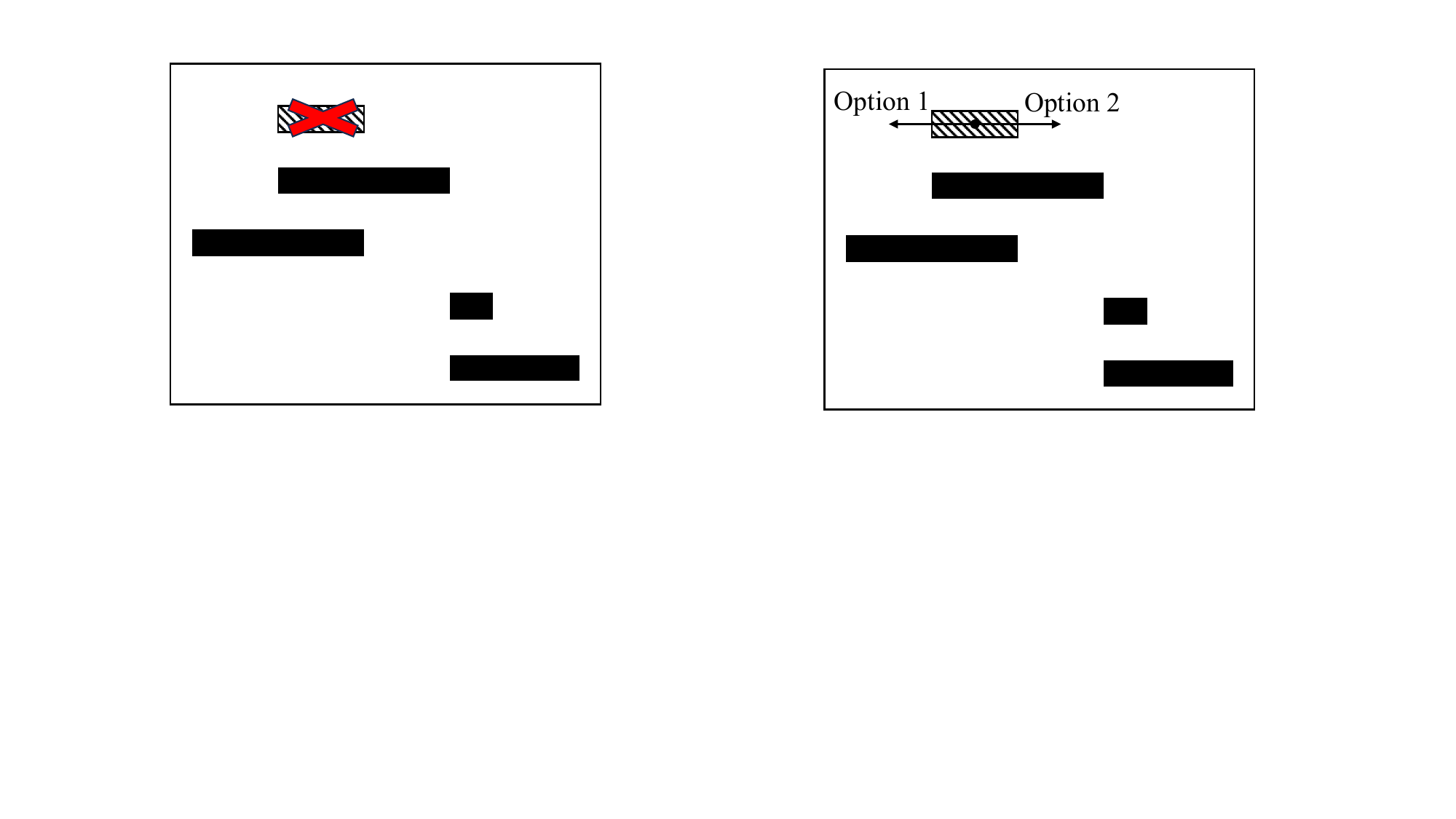};
  \end{axis}
\end{tikzpicture}
 \captionsetup{skip=-3pt}
   \caption{}
   \label{fig:maintenanceflex}
\end{subfigure}
\caption{Gantt chart of scheduled maintenance assets $1$-$4$ and a new request of asset $5$. In (a), the request does not fulfill physical requirements, and the TSO may decline. However, in (b), negotiating the starting time could resolve this physical infeasibility, e.g., proposing the user to shift maintenance of $5$ by two weeks earlier (opt 1) or later (opt 2). }%\vspace{-0.3em}}
\label{fig:maintenance}
\end{figure}

\normalsize
The function $h$ \textcolor{black}{in Alg. \ref{alg2}} solves the AC power flow for each time step $t \in \Omega^T$ where $P_t$ is added to the loads $P_D$ at the specific bus $b$ (we here call it also zip-code). The AC power flow outputs all voltage magnitudes $V_{m,t}$ at the bus $m$ and the line loadings $L_{l,t}$ at the line $l$ for all lines $l \in \Omega^L$ and buses $m \in \Omega^M$, respectively. Within the function $h$, the security of the connection is assessed using specified limits on voltage magnitudes and line loadings
%\vspace{-0.5em}

\begin{subequations}
\small
    \begin{align}
\bigwedge\limits_{m \in \Omega^M} V^{\min} & \leq V_{m,\hat{t}} \leq V^{\max} \label{eq:ACpowerflow1} \\ 
\bigwedge\limits_{l \in \Omega^L} L^{\min} & \leq L_{l,\hat{t}} \leq L^{\max} \label{eq:ACpowerflow2} 
\end{align}
\end{subequations}
\normalsize
to provide a security label each time
%\vspace{-0.5em}
\begin{equation}
s_{\hat{t}} = \begin{cases*} \label{eq:security}
 1, & if \eqref{eq:ACpowerflow1} and \eqref{eq:ACpowerflow2} hold \\
  0,                    & otherwise. %\vspace{-0.5em}
\end{cases*}
\end{equation}
The electricity connection is feasible iff $ \bigwedge\limits_{\hat{t} \in \Omega^T} s_{\hat{t}} = 1$. The output of the function $h$ is that the connection is

%\vspace{-1em}
\small
\begin{equation}
\beta = \begin{cases*}
 \textit{'feasible.' } & if  $\bigwedge\limits_{\hat{t} \in \Omega^T} s_{\hat{t}} = 1$  \\
  \textit{'infeasible at times\, } \{\hat{t} \in \Omega^T \mid s_{\hat{t}} =0 \} \textit{.'} & otherwise,%\vspace{-0.5em}
\end{cases*}
\end{equation}

\normalsize
that is added to the chat context $\omega_i$ Sec. \ref{sec:funchat}, and subsequently used by the LLM to respond to the industrial consumer (see \eqref{eq:chatllm}). The chat can respond by offering an electricity contract to the industrial consumer if the connection is feasible, and otherwise, the chat continues to interact with the consumer, providing feedback on the infeasible times to trigger changes in the connection request. 

%\vspace{-1.1em}
\subsection{Outage management in high-voltage grids}
%\vspace{-0.25em}

\begin{figure}
\centering
\includegraphics[width=0.3\textwidth]{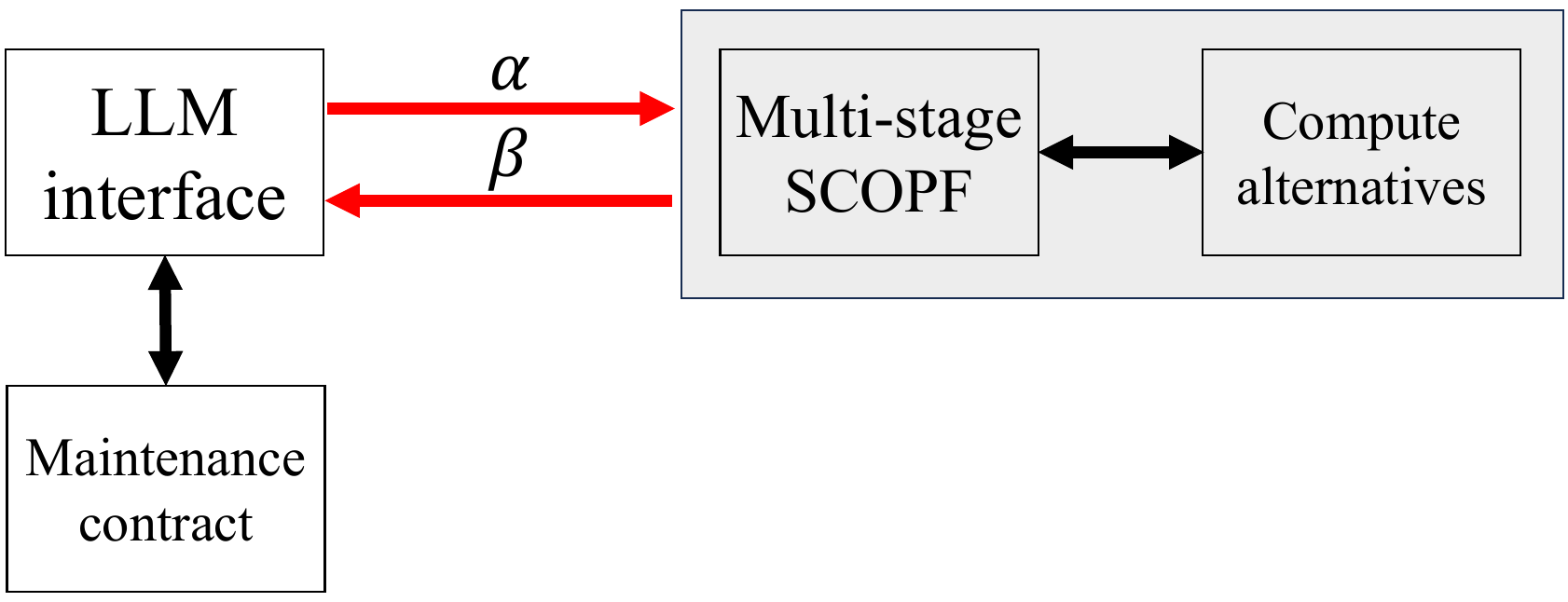}
   \caption{The maintenance planning approach. The red arrows show the exchanged parameters $\alpha$ and $\beta$.} %\vspace{-0.3em}}
   \label{fig:approachSCOPF}
\end{figure}
Maintenance is required as the transmission equipment ages and fails. The transmission system operator (TSO) manages scheduling these outages, \textcolor{black}{taking into account its} equipment and the generators connected to the grid. For example, in the EirGrid Grid Code~\cite{Eir25}, all planned outages must be submitted with sufficient notice. An asset owner sends a request to the TSO to disconnect their generator asset for a specified duration. As shown in Fig. \ref{fig:maintenancefixed}, TSOs can currently accept or decline requests. Upon decline, the asset owner may request a different maintenance time. These requests may accumulate at the TSO and take \textcolor{black}{months to plan (the second technical issue in Sec. \ref{sec:prob})}. For example, the National Grid ESO, UK, receives approximately $6000$ requests per year that are increasing and accumulating. This long planning horizon and an increase in requests require new tools to process these frequent requests more efficiently. Fig. \ref{fig:maintenanceflex} shows the proposed process in which the asset owner is presented, in near-realtime, with alternative options. In this process, the already planned maintenance of generators $g$ is $\tau_{gt} =\{0,1\} \quad \forall g \in \Omega^G$, showing the availability of the generator $g$ on day $t$. $\tau_{g,t} = 1$ indicates available, otherwise, $\tau_{gt} =0$, unavailable. Fig. \ref{fig:approachSCOPF} presents how the user, the asset owner, interacts with the LLM assessing the feasibility of the maintenance (outage) using a multi-stage security-constrained optimal power flows (SCOPF). The minimum information required for this tool is three scalar parameters $\alpha = [t^0,\Delta t, \hat{g}]$, the start day of the required outage $t^0$, the duration $\Delta t$, and the generator asset number $\hat{g}$. Subsequently, the requested outage of this generator $\tau_{\hat{g},\hat{t}} =0, \quad \forall \hat{t} \in \Omega^T$ is assessed for $\Omega^T = \{t^0, t^0+1, \dots, t^0 + \Delta t\}$. The SCOPFs
%\setlength{\tabcolsep}{1pt}

%\vspace{-1em} % tighten space below
\begin{subequations}\label{eq:SCOPF}
\small
    \begin{align}
    & \min_{P_{\text{G}},s_{\hat{t}}}  && \mathop{\mathlarger{\sum}} \limits _{g \in \Omega^\text{G}} \gamma_g P_{\text{G}g} + \lambda s_{\hat{t}} \\
    &  \text{s.t.} 
    && \tau_{g\hat{t}} P^{\min}_{\text{G}g} \leq P_{\text{G}g} \leq \tau_{g\hat{t}} P_{\text{G}g}^{\max} &&& \forall g \in \Omega^{\text{G}} \\
    & && \mathcal{G}(P_{\text{G}},P_{\text{D}\hat{t}}) \leq s_{\hat{t}}, 
        \end{align}
        \end{subequations}
\normalsize
may be solved for each time $\hat{t} \in \Omega^T$ to assess the physical feasibility. $\mathcal{G}(P_{\text{G}},P_{\text{D}\hat{t}})$ considers all node balances, power flow equations, N-$1$ security constraints and the power loading condition $P_{\text{D}\hat{t}}$. $P^{\min}_{\text{G}}$ and $P^{\max}_{\text{G}}$ are the lower and upper bounds of generator power levels, respectively, and $\gamma_g$ \textcolor{black}{is} the cost of generator $g$. $s \geq 0$ are slack variables for each constraint added as penalty terms to the objective function, making each optimisation feasible. The user specifies a large scalar value $\lambda \gg \gamma_g \quad \forall g \in \Omega^\text{G}$ to effectively penalise. After the SCOPFs are solved for all $\hat{t}$, the user maintenance request is assessed as physical feasible iff %$\lVert s_{\hat{t}} \rVert =0 \, \forall \hat{t} \in \Omega^T$, 
%\vspace{-1em}
\begin{equation} \small
\bigwedge\limits_{\hat{t} \in \Omega^T} \lVert s_{\hat{t}} \rVert =0, %\vspace{-0.5em}
\end{equation}
\normalsize
otherwise infeasible. Alg. \ref{alg3} implements the function $h_{k}(\alpha_k) \rightarrow \beta_k$ for the outage management (grey part in Fig. \ref{fig:approachSCOPF}) that is accessible by the LLM and considers the optimisation \eqref{eq:SCOPF}. Alg. \ref{alg3} also computes an alternative if the initial user request is infeasible, e.g. an earlier starting time $t^0$ if it exists. This algorithm can easily be expanded further to consider multiple alternatives earlier, later, or at different times. 

\setlength{\textfloatsep}{0pt}
\begin{algorithm}[t]
\small
\caption{\small Outage management $h(\alpha)\rightarrow \beta$}
\DontPrintSemicolon
\KwInput{$\alpha= [ t^0, \Delta t, \hat{g}]$}
\KwOutput{$\beta$}
\KwInit{$\tau, \, P_D, \lambda, $}
$\Omega^T = \{t^0, t^0+1, \dots, t^0 + \Delta t\}$, \;
$\tau_{\hat{g}\hat{t}} =0 \quad \forall \hat{t} \in \Omega^T$ \;
\For{each $\hat{t} \in \Omega^T$} {
solve \eqref{eq:SCOPF} with input $P_{\text{D}\hat{t}}$, $\tau_{\hat{g},\hat{t}}$ and output $s_{\hat{t}}$
}
\eIf{$\bigwedge\limits_{\hat{t} \in \Omega^T} \lVert s_{\hat{t}} \rVert =0$}{$\beta = $\textit{'requested outage plan is possible.'}}
{
%\beta = $\textit{'the requested outage plan is not possible'}
\While{$\left(\bigwedge\limits_{\hat{t} \in \Omega^{T}} \lVert s_{\hat{t}} \rVert >0\right) \bigwedge \left(t^0 \geq 0\right) $} {
$t^0 = t^0 -1$, $\tau_{\hat{g}{t^0}} =0$, \;
$\Omega^{T} = \{t^0, t^0+1, \dots, t^0 + \Delta t\}$ \;
solve \eqref{eq:SCOPF} with input $P_{\text{D}{t^0}}$, $\tau_{\hat{g},t^0}$ and output $s_{{t^0}}$
}
%\textit{$\beta = \beta +$' but starting at $t^0$ is possible.}'

\eIf{$\bigwedge\limits_{\hat{t} \in \Omega^T} \lVert s_{\hat{t}} \rVert =0$}
{$\beta = $\textit{'requested outage plan is not possible, but starting at $t^0$ is possible.'}}
{$\beta =$\textit{'requested outage plan is not possible.' }}
}
return $\beta$ 
\label{alg3}
\end{algorithm}

% \begin{figure}
% %\centering
% \begin{subfigure}[t]{0.18\textwidth}
% \includegraphics[width=1\textwidth]{figures/SCOPF.pdf}
%    \caption{}
%    \label{fig:approachSCOPF}
% \end{subfigure} \hspace{0.05\textwidth}
% \begin{subfigure}[t]{0.18\textwidth} 
% \begin{tikzpicture}
% \begin{axis}[
%         axis on top,% ----
%         width=\textwidth,
%         scale only axis,
%         enlargelimits=false, 
%         ytick={1,2,3,4,5,6},  
%         yticklabels= {1,2,3,4,5,$n$},  
%         xtick={0,1,2,3,4,5,6,7,8,9},
%         ylabel={Asset number [ ]},
%         xlabel={Time horizon [weeks]},
%         %xlabel style={at={(rel axis cs:1.05,0.17)}},
%         xmin=-0.5,
%         xmax=9.5,
%         ymin=0.5,
%         ymax=6.8,
%         %axis lines=middle,
%         %y label style={at={(axis description cs:0.2,.5)},anchor=south}, 
%         ]        
% 	\addplot graphics[xmin=-0.5,xmax=9.5,ymin=0.5,ymax=6.8] {figures/bruteforce.pdf};
%   \end{axis}
% \end{tikzpicture}
%    \caption{}
%    \label{fig:maxinformation}
% \end{subfigure}
% \caption{The maintenance planning approach (a) and possible privacy implications in (b). In (a), the red arrows show the exchanged parameters $\alpha$ and $\beta$. In (b), a brute force attack on the system that varies $\alpha$ to learn about the power network would maximally reveal information about the maintenance possibilities of each generator through $\beta$, for example, when assets $5$ to $n$ can be switched off. %\vspace{-1em}}
% \label{fig:privacy}
% \end{figure}

%\vspace{-1em}
\section{Case study}\label{sec:casestudy}
%\vspace{-0.5em}
\subsection{Test systems and assumptions}
\begin{figure}
%\centering
\begin{subfigure}[t]{0.15\textwidth}
\begin{tikzpicture}
\begin{axis}[
        axis on top,% ----
        width=\textwidth,
        scale only axis,
        enlargelimits=false, 
        ytick={2,4,6,8,10,12},     
        xtick={0,4,8,12,16,20,24},
        ylabel={Energy [kWh]},
        xlabel={Time [h]},
        xmin=0,
        xmax=24,
        ymin=2,
        ymax=13,
        ]        
	\addplot graphics[xmin=0,xmax=24,ymin=2,ymax=12] {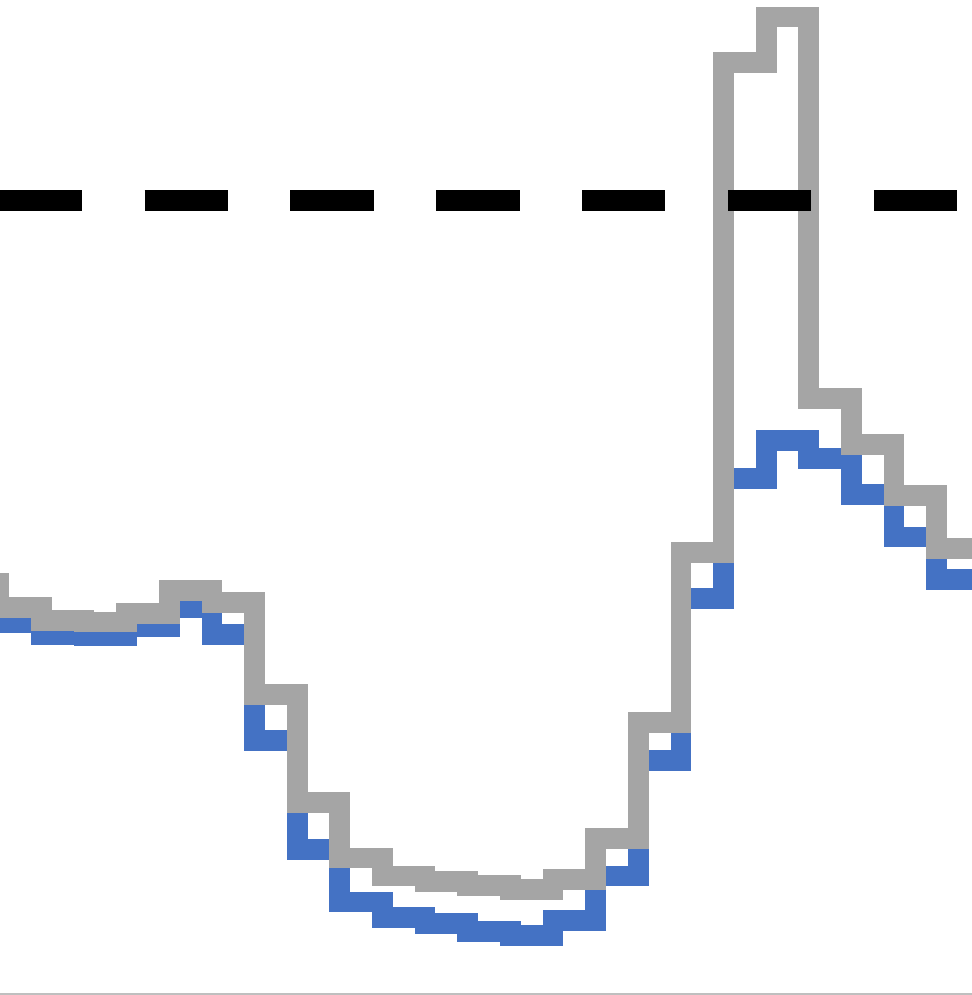};
  \end{axis}
\end{tikzpicture}
    \captionsetup{skip=-3pt}
   \caption{}
   \label{fig:initialrequest}
\end{subfigure} \hspace{0.1\textwidth}
\begin{subfigure}[t]{0.15\textwidth} 
\begin{tikzpicture}
\begin{axis}[
        axis on top,% ----
        width=\textwidth,
        scale only axis,
        enlargelimits=false, 
        ytick={2,4,6,8,10,12},     
        xtick={0,4,8,12,16,20,24},
        ylabel={Energy [kWh]},
        xlabel={Time [h]},
        xmin=0,
        xmax=24,
        ymin=2,
        ymax=13,
        ]        
	\addplot graphics[xmin=0,xmax=24,ymin=2,ymax=12] {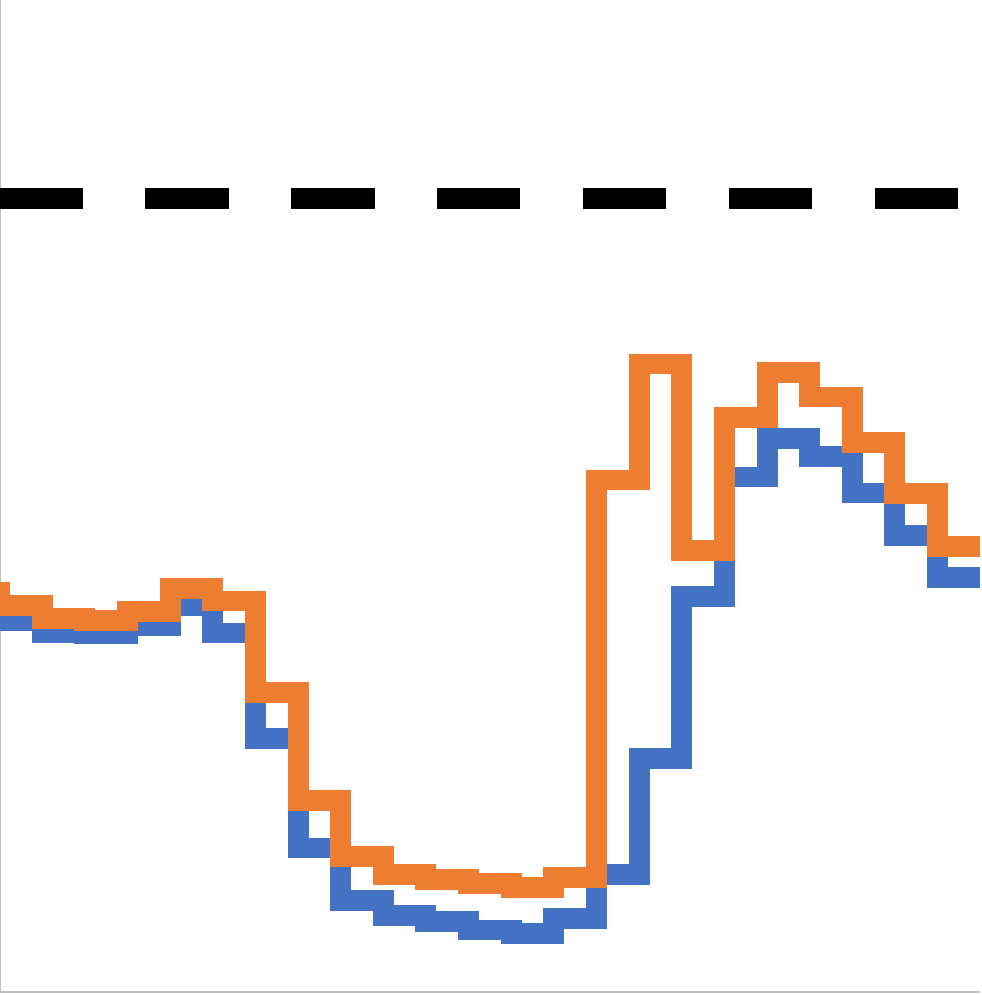};
  \end{axis}
\end{tikzpicture}
    \captionsetup{skip=-3pt}
   \caption{}
   \label{fig:offertouser}
\end{subfigure}
\caption{The expected power profile in this neighbourhood (black). (a) An initial user request with EV charging times at 17:00 would increase the total power profile in this neighbourhood (grey) and exceed the physical constraints of the low-voltage grid (dotted black). (b) is the adjusted total power profile (orange) if the charging starts at 15:00.}
\label{fig:flexibleshifting}
\end{figure}

The studies investigate the performance of the interactions with the chat and the use cases in Sec. \ref{sec:usecases}. \textcolor{black}{As all physical feasibility checks rely on established power system tools, performance comparisons focus on interaction robustness, correctness, and system-level utilisation rather than solver runtime.} In the first use case, connecting residential to the low-voltage system, we assumed a winter workday without a heat pump, aggregated hourly from \cite{Sch22}. We used a scaled $2019$ duck curve shown as the black line in Fig. \ref{fig:flexibleshifting} \cite{IEA19} for $\Omega^T = \{1, 2, \dots, 24\}$ hours. We assume a resident needs to charge daily $3.6\times1.15 = 4.14$kW for two hours ($1.15$ is a safety margin). A preferred charging time is $17$-$18$h. However, some neighbours may be flexible in shifting the charging to other times. We assume shared preference ${\rho}^{\text{shared}} _t = 0 \quad\forall t \in \{1 \dots 16, 19 \dots 24\} $, and ${\rho}^{\text{shared}}_t = 0.5 \quad\forall t \in \{17,18\}$. We sample preference noise $\eta_{u} \sim \text{Uniform}(0, 1)^{|\Omega^T|} \quad \text{for } u = 1, \dots, 15$ residents. We sample charging times from the probabilities %\vspace{-0.5em}
\begin{equation}
    \rho_u = (1 - \varepsilon)\, {\rho}^{\text{shared}} + \varepsilon\, \frac{\eta_{u}}{\sum_{i} \eta_i} %\vspace{-0.5em}
\end{equation}
either homogenous $\varepsilon=0$ or heterogenous charging preferences $\varepsilon=1$. For each resident two charging preferences were sampled $T_u \subset \Omega^T$ with $|T_u| = 2$ from $\rho_u$ without replacement. 

The second use case on improving contracting with SMEs was implemented on the $179$-bus Oberrhein MV grid applying the pandapower package \cite{Thu18}. The network is shown in Fig. \ref{fig:usecaseMV}, considering higher loading levels than the default network. The mean active loading level across all buses is shown in Fig. \ref{fig:pfoberrhein}. The (spatial-) correlation between loads is from the default network data. The active power levels of generators are accordingly increased. The default AC power flow was used, and the security was assessed with $V^{\min} = 0.975$, $V^{\max} = 1.03$, $L^{\min} = 0\%$ and $L^{\max} = 60\%$. Running the AC power flow shows the network is secure, only slightly below the voltage limit at at least one bus (see Fig. \ref{fig:pfoberrhein}). The chat can add an active load to one bus, and depending on where and what power level is added, it may make the system insecure. For example, Fig. \ref{fig:usecaseMV} shows all locations where adding a single, constant $P_t = 2\,\mathrm{MW} \quad \forall t \in \Omega^T$ load at all times leads to insecurity.

The third use case on outage management considered the IEEE $118$-bus system that has $18$ generators. A time horizon of $365$ days was considered. For each day, $10$ wind power plants and load data from the real French transmission system were used \cite{Sun18}. $10$ transmission lines were uniformly, randomly sampled as N-$1$ security constraints (shown in Fig. \ref{fig:118bus}). The load levels were scaled by $0.5$ to receive N-$1$ secure generator dispatches when solving $365$ SCOPFs for each day. Then, $50$\% of random generators were assumed with planned maintenance starting at a random week of the year for a duration of $4$ weeks. GUROBI 12.0.1 was used to solve the SCOPFs. Running Alg. \ref{alg3}, in the chat called 'OutagePlanner', solves $365$ SCOPFs and takes around $1$min in total on a standard laptop. 

The chats and LLMs were developed within the Wolfram Alpha Chatbook environment, which provides an API for LLM models and Python 3.11.8. \textcolor{black}{The code is available at \cite{llmcontracts}. Functional programs are implemented as tool calls matching $h(\alpha) \rightarrow \beta$ with instruction metadata $h^{inst}$
\eqref{eq:func}. Personas act as 
\eqref{eq:receiveorder}, determining which functions $\mathcal{H}_i \subset \mathcal{H}$ may run. Tool access is explicitly assigned by the developer of the system, realising the selection $\mathcal{H}$ so that each persona accesses only their specified set of functions.} Wolfram Mathematica 14.0 and 14.2 were used \cite{Mathematica}. The LLM models were 'Wolfram', 'OpenAI-$4$', 'OpenAI-$4$ Turbo' and 'OpenAI-$4$ Turbo Preview'.

%\vspace{-1em}
\subsection{\textcolor{black}{Improving grid efficiency by} customising electricity contracts for residents}
\begin{figure}
\scriptsize
%\captionsetup{format=plain}
\begin{assistantbubble}
\textbf{Energy supplier:} Hello! How can I assist you today? Are you looking to estimate your electricity consumption or need help with something else?
\end{assistantbubble}

\begin{userbubble}
\textbf{User:} yes
\end{userbubble}

\begin{assistantbubble}
\textbf{Energy supplier:} Great! Let's start by estimating your electricity consumption. I'll need to ask you a few questions to get a better idea of your usage. 
\begin{enumerate}
\item How many hours per day do you use your washing machine?
\item How long do you use your dishwasher each day?
\item How many hours is your TV on daily?
\item How many hours do you use your computer each day?
\item How long are your lights typically on each day?
\item Do you have an electric vehicle? If so, how many hours do you charge it daily, and what time do you usually start charging it?
\end{enumerate}
Feel free to provide the details, and we'll proceed from there!
\end{assistantbubble}

\begin{userbubble}
\textbf{User:}  $3$h washing, $2$h dish, compute none, lights $8$h, EV $1$h at $18:00$
\end{userbubble}

\begin{assistantbubble}
\textbf{Enery Supplier:} 
Thank you for providing the details. Let's summarize your input: 
\begin{itemize}
\item Washing machine: 3 hours per day
\item Dishwasher: 2 hours per day
\item Computer: 0 hours per day
\item Lights: 8 hours per day
\item Electric Vehicle (EV): 1 hour per day starting at 18:00
\end{itemize}
I will now calculate an estimate of your average electricity consumption per hour using this data. \vspace{-1.5em}
\begin{center}
\begin{tcolorbox}[
    colback=green!10,
    colframe=green!60!black,
    arc=1mm,
    boxrule=0.8pt,
    left=2mm, right=2mm,
    top=0mm, bottom=0mm,
    enhanced,
    width=4.5cm,
    halign=center,
    valign=center,
    box align=base,
    boxsep=0.5mm,      % tight internal padding
    before skip=1pt,
    after skip=1pt,
    fontupper={\scriptsize\setlength{\baselineskip}{10pt}}  % smaller text and tighter line height
]
\tikzwrench~Used ElectricityConsumption~\tikzwrench
\end{tcolorbox}
\end{center}
\vspace{-0.5em}
I've calculated your estimated electricity consumption per hour based on the information you provided. Here's the breakdown of your daily electricity profile (in kWh) for each hour of the day:

\vspace{-1.5em}
\begin{multicols}{2}
\begin{enumerate}
\item 00:00 - 01:00: 0.1619 kWh
\item 01:00 - 02:00: 0.1198 kWh
\item 02:00 - 03:00: 0.1099 kWh
\item 03:00 - 04:00: 0.1069 kWh
\item 04:00 - 05:00: 0.1087 kWh
\item 05:00 - 06:00: 0.1302 kWh
\item 06:00 - 07:00: 0.2502 kWh
\item 07:00 - 08:00: 0.3595 kWh
\item 08:00 - 09:00: 0.3708 kWh
\item 09:00 - 10:00: 0.3422 kWh
\item 10:00 - 11:00: 0.3229 kWh
\item 11:00 - 12:00: 0.3287 kWh
\item 12:00 - 13:00: 0.3631 kWh
\item 13:00 - 14:00: 0.3597 kWh
\item 14:00 - 15:00: 0.3204 kWh
\item 15:00 - 16:00: 0.2912 kWh
\item 16:00 - 17:00: 0.2923 kWh
\item 17:00 - 18:00: 0.3574 kWh
\item 18:00 - 19:00: 4.0616 kWh (includes charging your EV)
\item 19:00 - 20:00: 0.5196 kWh
\item 20:00 - 21:00: 0.4673 kWh
\item 21:00 - 22:00: 0.3916 kWh
\item 22:00 - 23:00: 0.3241 kWh
\item 23:00 - 00:00: 0.2400 kWh
\end{enumerate}
\end{multicols}
\vspace{-1.5em}
Now, I will check if this profile is feasible with the power system network constraints and develop a suitable energy supply contract for you. 
\vspace{-0.5em}
\begin{center}
\begin{tcolorbox}[
    colback=green!10,
    colframe=green!60!black,
    arc=1mm,
    boxrule=0.8pt,
    left=2mm, right=2mm,
    top=0mm, bottom=0mm,
    enhanced,
    width=4cm,
    halign=center,
    valign=center,
    box align=base,
    boxsep=0.5mm,      % tight internal padding
    before skip=1pt,
    after skip=1pt,
    fontupper={\scriptsize\setlength{\baselineskip}{10pt}}  % smaller text and tighter line height
]
\tikzwrench~Used DevelopContract~\tikzwrench
\end{tcolorbox}
\end{center}
\vspace{-0.5em}
It seems that the current electricity consumption profile is not feasible with the power system network constraints due to the high demand when charging your electric vehicle at 18:00. However, alternative charging times are suggested to make it feasible. You can consider charging your EV during the following time slots instead: 00:00 - 17:00. Would you like to adjust the EV charging time to one of these alternative time slots? If so, please let me know the preferred time, and I will recalculate the consumption and contract details for you.
\end{assistantbubble}

\begin{userbubble}
\textbf{User:} yes, it would be great if you could adjust for 16:00
\end{userbubble}

\begin{assistantbubble}
\textbf{Energy Supplier:} Great! Let's adjust the EV charging time to start at 16:00 and see if a feasible contract can be developed. I'll recalculate the electricity consumption profile with this new charging time.  \vspace{-0.5em}
\begin{center}
\begin{tcolorbox}[
    colback=green!10,
    colframe=green!60!black,
    arc=1mm,
    boxrule=0.8pt,
    left=2mm, right=2mm,
    top=0mm, bottom=0mm,
    enhanced,
    width=4.5cm,
    halign=center,
    valign=center,
    box align=base,
    boxsep=0.5mm,      % tight internal padding
    before skip=1pt,
    after skip=1pt,
    fontupper={\scriptsize\setlength{\baselineskip}{10pt}}  % smaller text and tighter line height
]
\tikzwrench~Used ElectricityConsumption~\tikzwrench
\end{tcolorbox}
\end{center}
\vspace{-0.5em}
I've recalculated your estimated electricity consumption profile with the EV charging time adjusted to start at 16:00. Here's the updated daily electricity profile (in kWh) for each hour of the day:
\vspace{-1.5em}
\begin{multicols}{2}
\begin{enumerate}
\item 00:00 - 01:00: 0.1619 kWh
\item01:00 - 02:00: 0.1198 kWh
\item 02:00 - 03:00: 0.1099 kWh
\item 03:00 - 04:00: 0.1069 kWh
\item 04:00 - 05:00: 0.1087 kWh
\item 05:00 - 06:00: 0.1302 kWh
\item 06:00 - 07:00: 0.2502 kWh
\item 07:00 - 08:00: 0.3595 kWh
\item 08:00 - 09:00: 0.3708 kWh
\item 09:00 - 10:00: 0.3422 kWh
\item 10:00 - 11:00: 0.3229 kWh
\item 11:00 - 12:00: 0.3287 kWh
\item 12:00 - 13:00: 0.3631 kWh
\item 13:00 - 14:00: 0.3597 kWh
\item 14:00 - 15:00: 0.3204 kWh
\item 15:00 - 16:00: 0.2912 kWh
\item 16:00 - 17:00: 3.8923 kWh (includes charging your EV)
\item 17:00 - 18:00: 0.3574 kWh
\item 18:00 - 19:00: 0.4616 kWh
\item 19:00 - 20:00: 0.5196 kWh
\item 20:00 - 21:00: 0.4673 kWh
\item 21:00 - 22:00: 0.3916 kWh
\item 22:00 - 23:00: 0.3241 kWh
\item 23:00 - 00:00: 0.2400 kWh
\end{enumerate}
\end{multicols}
\vspace{-1.5em}
Now, I'll check if this updated profile is feasible and develop a suitable energy supply contract for you. \vspace{-1.5em}
\begin{center}
\begin{tcolorbox}[
    colback=green!10,
    colframe=green!60!black,
    arc=1mm,
    boxrule=0.8pt,
    left=2mm, right=2mm,
    top=0mm, bottom=0mm,
    enhanced,
    width=4cm,
    halign=center,
    valign=center,
    box align=base,
    boxsep=0.5mm,      % tight internal padding
    before skip=1pt,
    after skip=1pt,
    fontupper={\scriptsize\setlength{\baselineskip}{10pt}}  % smaller text and tighter line height
]
\tikzwrench~Used DevelopContract~\tikzwrench
\end{tcolorbox}
\end{center}
\vspace{-0.5em}
Good news! The updated electricity consumption profile with the EV charging time adjusted to start at 16:00 is feasible. A suitable energy supply contract can be offered to you with the following details: 
\begin{itemize}
\item \textbf{Base Level Consumption}: 0.6 kWh
\item \textbf{Flexible Level During Charging Hours}: 4.8 kWh
\end{itemize}
This contract will accommodate your electricity needs efficiently, including the charging of your electric vehicle. If you have any questions or need further assistance, feel free to ask!
\end{assistantbubble}

\caption{Example chat between an energy supplier and a residential user. }
\label{fig:LVchat}
\end{figure}
%This study estimates the value of using customised contracts over standard block contracts. 
\textcolor{black}{Standard contracts are based on the assumption of homogeneity at the user, leading to grid inefficiency. This study investigates the potential efficiency increase when using customised over standard contracts for the residential use case (Sec. \ref{sec:reschat}). The study assumes the following for the proposed system:} the residential user sends an initial request $r_t \quad \forall t \in \Omega^T$ at times $\Omega^T=\{1,2,\cdots, 24\}$ to an energy supplier via the chat (example in Fig. \ref{fig:LVchat}). The chat-system adjusts the profile $r_t \quad \forall t \in \Omega^T$ by shifting the flexible part using functions $h_1$ and $h_2$ introduced in Sec. \ref{sec:reschat}. %A conventional block contract may assume $P = \varepsilon \times \max \{r_t | t \in \Omega^T\}$ where $\varepsilon=0.15$ may be a safety margin. 
The proposed customised contracts can have off-peak and on-peak times (as shown in Fig. \ref{fig:contract}) or customised price levels leading to a potential lower energy profile shown in Fig. \ref{fig:flexibleshifting}. The aggregated neighbourhood profile is the black curve. In a neighbourhood with many heterogeneous power profiles, this can lead to a 'flatter' consumption profile. 

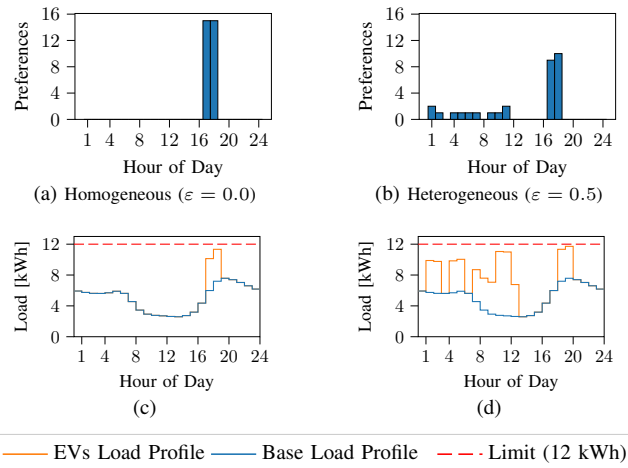
\begin{figure}
\centering

% Top row
\begin{subfigure}[t]{0.2\textwidth}
  \centering
  \adjustbox{width=\linewidth}{% This file was created with tikzplotlib v0.10.1.
\begin{tikzpicture}

\definecolor{darkgray176}{RGB}{176,176,176}
\definecolor{steelblue31119180}{RGB}{31,119,180}

\begin{axis}[height=0.08\textheight,width=\textwidth,scale only axis,
tick align=outside,
tick pos=left,
x grid style={darkgray176},
xlabel={Hour of Day},
xmin=-0.7, xmax=25.7,
xtick style={color=black},
xtick={1,4,8,12,16,20,24},
xticklabels={
$1$,
  $4$,
  $8$,
  $12$,
  $16$,
  $20$,
  $24$
},
y grid style={darkgray176},
ylabel={Preferences},
ymin=0, ymax=16.1,
ytick style={color=black},
ytick={0,4,8,12,16},
yticklabels={
  ${0}$,
  ${4}$,
  ${8}$,
  ${12}$,
  ${16}$
}
]
\draw[draw=black,fill=steelblue31119180] (axis cs:0.5,0) rectangle (axis cs:1.5,0);
\draw[draw=black,fill=steelblue31119180] (axis cs:1.5,0) rectangle (axis cs:2.5,0);
\draw[draw=black,fill=steelblue31119180] (axis cs:2.5,0) rectangle (axis cs:3.5,0);
\draw[draw=black,fill=steelblue31119180] (axis cs:3.5,0) rectangle (axis cs:4.5,0);
\draw[draw=black,fill=steelblue31119180] (axis cs:4.5,0) rectangle (axis cs:5.5,0);
\draw[draw=black,fill=steelblue31119180] (axis cs:5.5,0) rectangle (axis cs:6.5,0);
\draw[draw=black,fill=steelblue31119180] (axis cs:6.5,0) rectangle (axis cs:7.5,0);
\draw[draw=black,fill=steelblue31119180] (axis cs:7.5,0) rectangle (axis cs:8.5,0);
\draw[draw=black,fill=steelblue31119180] (axis cs:8.5,0) rectangle (axis cs:9.5,0);
\draw[draw=black,fill=steelblue31119180] (axis cs:9.5,0) rectangle (axis cs:10.5,0);
\draw[draw=black,fill=steelblue31119180] (axis cs:10.5,0) rectangle (axis cs:11.5,0);
\draw[draw=black,fill=steelblue31119180] (axis cs:11.5,0) rectangle (axis cs:12.5,0);
\draw[draw=black,fill=steelblue31119180] (axis cs:12.5,0) rectangle (axis cs:13.5,0);
\draw[draw=black,fill=steelblue31119180] (axis cs:13.5,0) rectangle (axis cs:14.5,0);
\draw[draw=black,fill=steelblue31119180] (axis cs:14.5,0) rectangle (axis cs:15.5,0);
\draw[draw=black,fill=steelblue31119180] (axis cs:15.5,0) rectangle (axis cs:16.5,0);
\draw[draw=black,fill=steelblue31119180] (axis cs:16.5,0) rectangle (axis cs:17.5,15);
\draw[draw=black,fill=steelblue31119180] (axis cs:17.5,0) rectangle (axis cs:18.5,15);
\draw[draw=black,fill=steelblue31119180] (axis cs:18.5,0) rectangle (axis cs:19.5,0);
\draw[draw=black,fill=steelblue31119180] (axis cs:19.5,0) rectangle (axis cs:20.5,0);
\draw[draw=black,fill=steelblue31119180] (axis cs:20.5,0) rectangle (axis cs:21.5,0);
\draw[draw=black,fill=steelblue31119180] (axis cs:21.5,0) rectangle (axis cs:22.5,0);
\draw[draw=black,fill=steelblue31119180] (axis cs:22.5,0) rectangle (axis cs:23.5,0);
\draw[draw=black,fill=steelblue31119180] (axis cs:23.5,0) rectangle (axis cs:24.5,0);
\end{axis}
\end{tikzpicture}}
  \captionsetup{skip=0pt, justification=centering}
  \caption{\scriptsize Homogeneous ($\varepsilon = 0.0$)}
  \label{fig:histhet0}
\end{subfigure}
\hspace{2em}
\begin{subfigure}[t]{0.2\textwidth}
  \centering
  \adjustbox{width=\linewidth}{% This file was created with tikzplotlib v0.10.1.
\begin{tikzpicture}

\definecolor{darkgray176}{RGB}{176,176,176}
\definecolor{steelblue31119180}{RGB}{31,119,180}

\begin{axis}[height=0.08\textheight,width=\textwidth,scale only axis,
tick align=outside,
tick pos=left,
x grid style={darkgray176},
xlabel={Hour of Day},
xmin=-0.7, xmax=25.7,
xtick style={color=black},
xtick={1,4,8,12,16,20,24},
xticklabels={
$1$,
  $4$,
  $8$,
  $12$,
  $16$,
  $20$,
  $24$
},
y grid style={darkgray176},
ylabel={Preferences},
ymin=0, ymax=16.1,
ytick style={color=black},
ytick={0,4,8,12,16},
yticklabels={
  ${0}$,
  ${4}$,
  ${8}$,
  ${12}$,
  ${16}$
}
]
\draw[draw=black,fill=steelblue31119180] (axis cs:0.5,0) rectangle (axis cs:1.5,2);
\draw[draw=black,fill=steelblue31119180] (axis cs:1.5,0) rectangle (axis cs:2.5,1);
\draw[draw=black,fill=steelblue31119180] (axis cs:2.5,0) rectangle (axis cs:3.5,0);
\draw[draw=black,fill=steelblue31119180] (axis cs:3.5,0) rectangle (axis cs:4.5,1);
\draw[draw=black,fill=steelblue31119180] (axis cs:4.5,0) rectangle (axis cs:5.5,1);
\draw[draw=black,fill=steelblue31119180] (axis cs:5.5,0) rectangle (axis cs:6.5,1);
\draw[draw=black,fill=steelblue31119180] (axis cs:6.5,0) rectangle (axis cs:7.5,1);
\draw[draw=black,fill=steelblue31119180] (axis cs:7.5,0) rectangle (axis cs:8.5,0);
\draw[draw=black,fill=steelblue31119180] (axis cs:8.5,0) rectangle (axis cs:9.5,1);
\draw[draw=black,fill=steelblue31119180] (axis cs:9.5,0) rectangle (axis cs:10.5,1);
\draw[draw=black,fill=steelblue31119180] (axis cs:10.5,0) rectangle (axis cs:11.5,2);
\draw[draw=black,fill=steelblue31119180] (axis cs:11.5,0) rectangle (axis cs:12.5,0);
\draw[draw=black,fill=steelblue31119180] (axis cs:12.5,0) rectangle (axis cs:13.5,0);
\draw[draw=black,fill=steelblue31119180] (axis cs:13.5,0) rectangle (axis cs:14.5,0);
\draw[draw=black,fill=steelblue31119180] (axis cs:14.5,0) rectangle (axis cs:15.5,0);
\draw[draw=black,fill=steelblue31119180] (axis cs:15.5,0) rectangle (axis cs:16.5,0);
\draw[draw=black,fill=steelblue31119180] (axis cs:16.5,0) rectangle (axis cs:17.5,9);
\draw[draw=black,fill=steelblue31119180] (axis cs:17.5,0) rectangle (axis cs:18.5,10);
\draw[draw=black,fill=steelblue31119180] (axis cs:18.5,0) rectangle (axis cs:19.5,0);
\draw[draw=black,fill=steelblue31119180] (axis cs:19.5,0) rectangle (axis cs:20.5,0);
\draw[draw=black,fill=steelblue31119180] (axis cs:20.5,0) rectangle (axis cs:21.5,0);
\draw[draw=black,fill=steelblue31119180] (axis cs:21.5,0) rectangle (axis cs:22.5,0);
\draw[draw=black,fill=steelblue31119180] (axis cs:22.5,0) rectangle (axis cs:23.5,0);
\draw[draw=black,fill=steelblue31119180] (axis cs:23.5,0) rectangle (axis cs:24.5,0);
\end{axis}
\end{tikzpicture}}
  \captionsetup{skip=0pt, justification=centering}
  \caption{\scriptsize Heterogeneous ($\varepsilon = 0.5$)}
  \label{fig:histhet05}
\end{subfigure}

%\vspace{1em}

% Bottom row
\begin{subfigure}[t]{0.2\textwidth}
  \centering
  \adjustbox{width=\linewidth}{% This file was created with tikzplotlib v0.10.1.
\begin{tikzpicture}

\definecolor{darkgray176}{RGB}{176,176,176}
\definecolor{darkorange25512714}{RGB}{255,127,14}
\definecolor{lightgray204}{RGB}{204,204,204}
\definecolor{steelblue31119180}{RGB}{31,119,180}

\begin{axis}[height=0.08\textheight,width=\textwidth,scale only axis,
legend cell align={left},
legend style={at={(0.5,-0.2)}, font=\footnotesize,
anchor=north,
 /tikz/column 4/.style={column sep=5pt,},
  fill opacity=0.8,
  draw opacity=1,
  text opacity=1,
  draw=lightgray204
},
legend columns=4, 
legend cell align={left},
legend to name={mylegend},
% legend cell align={left},
% legend style={
%   fill opacity=0.8,
%   draw opacity=1,
%   text opacity=1,
%   at={(0.03,0.03)},
%   anchor=south west,
%   draw=lightgray204
% },
tick align=outside,
tick pos=left,
x grid style={darkgray176},
xlabel={Hour of Day},
xmin=0, xmax=24,
xtick style={color=black},
xtick={1,4,8,12,16,20,24},
xticklabels={
$1$,
  $4$,
  $8$,
  $12$,
  $16$,
  $20$,
  $24$
},
y grid style={darkgray176},
ylabel={Load [kWh]},
ymin=0, ymax=13,
ytick style={color=black},
ytick={0,4,8,12},
yticklabels={
  ${0}$,
  ${4}$,
  ${8}$,
  ${12}$
}
]
\addplot [semithick, darkorange25512714, const plot mark left]
table {%
0 5.925399521
1 5.741984523
2 5.622371131
3 5.610962102
4 5.697385619
5 5.897075954
6 5.620658048
7 4.555283772
8 3.452930057
9 2.925877861
10 2.769402738
11 2.708569715
12 2.625775705
13 2.58962188
14 2.740526059
15 3.186498979
16 4.354809723
17 10.128061814
18 11.334197641
19 7.578378703
20 7.396569432
21 7.033159709
22 6.605886341
23 6.178612972
24 6.178612972
};
\addlegendentry{EVs Load Profile}
\addplot [semithick, steelblue31119180, const plot mark left]
table {%
0 5.925399521
1 5.741984523
2 5.622371131
3 5.610962102
4 5.697385619
5 5.897075954
6 5.620658048
7 4.555283772
8 3.452930057
9 2.925877861
10 2.769402738
11 2.708569715
12 2.625775705
13 2.58962188
14 2.740526059
15 3.186498979
16 4.354809723
17 5.988061814
18 7.194197641
19 7.578378703
20 7.396569432
21 7.033159709
22 6.605886341
23 6.178612972
24 6.178612972
};
\addlegendentry{Base Load Profile}
\addplot [semithick, red, dash pattern=on 5.55pt off 2.4pt]
table {%
0 12
24 12
};
\addlegendentry{Limit (12 kWh)}
% \draw (axis cs:17.25,8.334197641) node[
%   scale=0.75,
%   anchor=base west,
%   text=black,
%   rotate=0.0
% ]{1 EVs};
\end{axis}

\end{tikzpicture}}
  \captionsetup{skip=0pt, justification=centering}
  \caption{}
  \label{fig:EVpower0}
\end{subfigure}
\hspace{2em}
\begin{subfigure}[t]{0.2\textwidth}
  \centering
  \adjustbox{width=\linewidth}{% This file was created with tikzplotlib v0.10.1.
\begin{tikzpicture}

\definecolor{darkgray176}{RGB}{176,176,176}
\definecolor{darkorange25512714}{RGB}{255,127,14}
\definecolor{lightgray204}{RGB}{204,204,204}
\definecolor{steelblue31119180}{RGB}{31,119,180}

\begin{axis}[height=0.08\textheight,width=\textwidth,scale only axis,
legend cell align={left},
legend style={
  fill opacity=0.8,
  draw opacity=1,
  text opacity=1,
  at={(0.03,0.03)},
  anchor=south west,
  draw=lightgray204
},
tick align=outside,
tick pos=left,
x grid style={darkgray176},
xlabel={Hour of Day},
xmin=0, xmax=24,
xtick style={color=black},
xtick={1,4,8,12,16,20,24},
xticklabels={
$1$,
  $4$,
  $8$,
  $12$,
  $16$,
  $20$,
  $24$
},
y grid style={darkgray176},
ylabel={Load [kWh]},
ymin=0, ymax=13,
ytick style={color=black},
ytick={0,4,8,12},
yticklabels={
  ${0}$,
  ${4}$,
  ${8}$,
  ${12}$
}
]
\addplot [semithick, darkorange25512714, const plot mark left]
table {%
0 5.925399521
1 9.881984523
2 9.762371131
3 5.610962102
4 9.837385619
5 10.037075954
6 5.620658048
7 8.695283772
8 7.592930057
9 7.065877861
10 11.049402738
11 10.988569715
12 6.765775705
13 2.58962188
14 2.740526059
15 3.186498979
16 4.354809723
17 5.988061814
18 11.334197641
19 11.718378703
20 7.396569432
21 7.033159709
22 6.605886341
23 6.178612972
24 6.178612972
};
%\addlegendentry{EVs Load Profile}
\addplot [semithick, steelblue31119180, const plot mark left]
table {%
0 5.925399521
1 5.741984523
2 5.622371131
3 5.610962102
4 5.697385619
5 5.897075954
6 5.620658048
7 4.555283772
8 3.452930057
9 2.925877861
10 2.769402738
11 2.708569715
12 2.625775705
13 2.58962188
14 2.740526059
15 3.186498979
16 4.354809723
17 5.988061814
18 7.194197641
19 7.578378703
20 7.396569432
21 7.033159709
22 6.605886341
23 6.178612972
24 6.178612972
};
%\addlegendentry{Base Load Profile}
\addplot [semithick, red, dash pattern=on 5.55pt off 2.4pt]
table {%
0 12
24 12
};
%\addlegendentry{Limit (12 kW)}
% \draw (axis cs:18.25,8.718378703) node[
%   scale=0.75,
%   anchor=base west,
%   text=black,
%   rotate=0.0
% ]{7 EVs};
\end{axis}

\end{tikzpicture}}
  \captionsetup{skip=0pt, justification=centering}
  \caption{}
  \label{fig:EVpower05}
\end{subfigure}

%\vspace{0.5em}

% Optional legend
\definecolor{darkgray176}{RGB}{176,176,176}
\definecolor{darkorange25512714}{RGB}{255,127,14}
\definecolor{lightgray204}{RGB}{204,204,204}
\definecolor{steelblack31119180}{RGB}{31,119,180}
\ref{mylegend}

\caption{Comparing user preferences and EV charging contracts under different heterogeneity levels. $15$ households with the charging preferences (a) can connect only $1$ EV (c). However, customising contracts exploiting heterogeneous preferences (b) allows $7$ EVs to connect (d).}
\label{fig:value}
%\vspace{-1.5em}
\end{figure}

\begin{figure}
 \centering
% This file was created with tikzplotlib v0.10.1.
\begin{tikzpicture}

\definecolor{darkgray176}{RGB}{176,176,176}
\definecolor{lightgray204}{RGB}{204,204,204}
\definecolor{steelblue31119180}{RGB}{31,119,180}

\begin{axis}[height=0.08\textheight,width=0.35\textwidth,
height = 3em,
scale only axis,
tick align=outside,
tick pos=left,
x grid style={darkgray176},
xlabel={Heterogeneity level $\varepsilon$},
xmin=-0.05, xmax=1.05,
xtick style={color=black},
xtick={-0.2,0,0.2,0.4,0.6,0.8,1,1.2},
xticklabels={
  ${−0.2}$,
  ${0.0}$,
  ${0.2}$,
  ${0.4}$,
  ${0.6}$,
  ${0.8}$,
  ${1.0}$,
  ${1.2}$
},
y grid style={darkgray176},
ylabel={EVs},
ymin=-0.1, ymax=10.555,
ytick style={color=black},
ytick={0,5,10},
yticklabels={
  ${0}$,
  ${5}$,
  ${10}$
}
]
\addplot [semithick, steelblue31119180, mark=*, mark size=3, mark options={solid}]
table {%
0 1
0.1 4.1
0.2 6.3
0.3 6.7
0.4 8.4
0.5 8
0.6 8.6
0.7 9.1
0.8 9.3
0.9 9.8
1 10.1
};
%\addlegendentry{EVs vs. Heterogeneity}
\end{axis}

\end{tikzpicture}
\caption{Average number of EVs that can connect depends on the heterogeneity in the charging time preferences ($\varepsilon = 0$ homogeneous, $\varepsilon = 1$ heterogeneous). }%\vspace{-1em}}
\label{fig:heterogenity}
\end{figure}
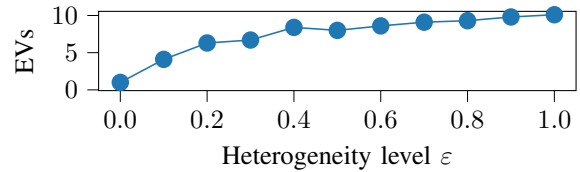

We investigated the impact of customising contracts in a neighbourhood of $15$ residents who have varying EV charging preferences. We applied a mixed integer linear program that maximised EV connections constrained by $12$ kWh maximal capacity for each hour for varying heterogeneity of charging time preferences $\varepsilon=\{0,0.1, \dots 1\}$, which we each repeat $10$ times. Fig. \ref{fig:value} shows that more EVs can be connected if the residents have varying preferences \textcolor{black}{and} individual customised contracts. In this example, $7$ times more EVs can be connected with the same grid capacity, as each resident received a different, preferred time slot to charge. Fig. \ref{fig:heterogenity} shows that already a low level of heterogeneity enables DSOs and energy suppliers to connect more EVs to the same low voltage system, as long as this heterogeneity can be exploited with customised contracts at large. \textcolor{black}{In other words, the grid can be used at a constantly higher level when customising capacity contracts over standard contracts, assigning each household the same capacity at all times.}

%\vspace{-1em}
\subsection{\textcolor{black}{Function calling with LLMs in residential contracting}}
\begin{figure}
\centering
\begin{tikzpicture}
\begin{axis}[
        axis on top,% ----
        width=0.14\textwidth,
        scale only axis,
        enlargelimits=false, 
        ytick={0,25, 50,75,100,125},     
        xtick={0,0.25,0.5,0.75,1},
        ylabel={Accuracy [\%]},
        xlabel={Temperature [ ]},
        xmin=0,
        xmax=1,
        ymin=50,
        ymax=125,
        ]        
	\addplot graphics[xmin=-0.05,xmax=1.05,ymin=-5,ymax=125] {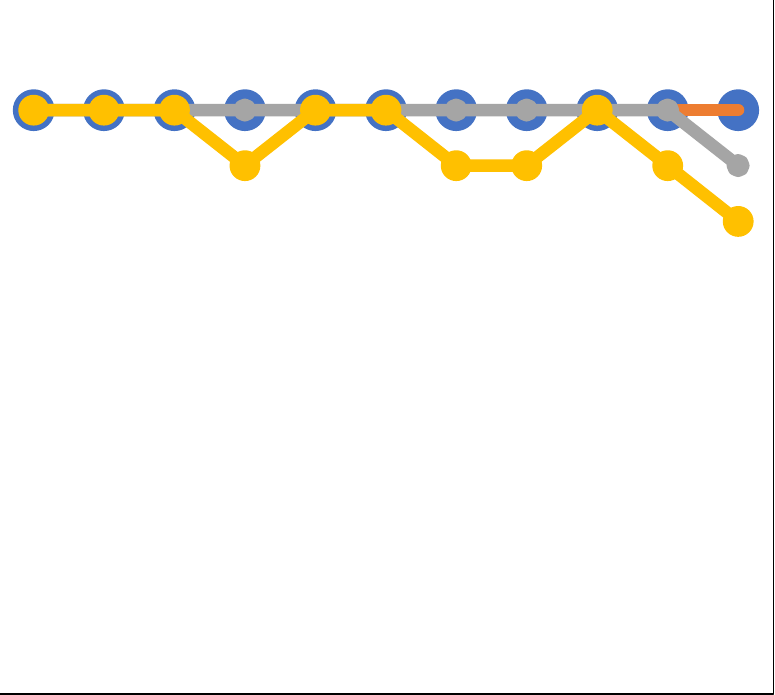};
  \end{axis}
\end{tikzpicture}
\caption{Performances of $g$ to correctly identifying parameters $\alpha$ from text $u_i$ (\textcolor[RGB]{68,114,196}{\rule[0.15em]{0.8em}{0.3em}}), of the functional program $h$ to correctly output $\beta$ (\textcolor[RGB]{237,125,49}{\rule[0.15em]{0.8em}{0.3em}}), of $v$ to pass information from program to program (\textcolor[RGB]{165,165,165}{\rule[0.15em]{0.8em}{0.3em}}) and of $f$ to include parameters $\beta$ in response $r_i$ (\textcolor[RGB]{255,192,0}{\rule[0.15em]{0.8em}{0.3em}}).}%\vspace{-1.5em}}
\label{fig:cstemp}
\end{figure}

\textcolor{black}{The proposed system combines LLM with analysis of the grid through function calling. This study investigates the accuracy of the tool calling (Sec. \ref{sec:funchat}), LLM settings and Alg. \ref{alg1} on the residential use case (Sec. \ref{sec:reschat})}. The paragraph $u^{0} = $\textit{'I use my washing machine $2$h and dishwasher $1.5$h every day. My TV is around $3$h on and computer $4$h. The lights are on $14$h a day. My EV charges $1$h every day starting at $10:00$. First, compute my electricity consumption by using the electricity consumption tool. Second, take my daily electricity consumption profile as input to check the physical feasibility with the physical feasible contract tool. Use this tool to provide me with a electricity supply contract. If the tool does not give a correct response, just simply try again and do not apologize.'} was tested $10$ times, where OpenAI's gpt-4-turbo-preview' has access to the two functions to estimate electricity consumption and check physical feasibility (Fig. \ref{fig:InterfaceLV}). The tuning temperature $\tau=0, 0.1, \dots 1$ was studied, so in total the paragraph was processed $110$ times. Fig. \ref{fig:cstemp} shows the accuracy of identifying parameters $\alpha$ from text, assessing the performance of $g$ in \eqref{eq:extraction}, and providing correct information $h$ in \eqref{eq:functionexecution} to the chat. \textcolor{black}{As all power-system computations are executed by the functional programs $h$, repeated requests with identical parameters $\alpha$ always return the same $\beta$. The variation observed in Fig. \ref{fig:cstemp}, therefore, originates solely from the extraction and chat process ($g$, $v$, ${f}$), rather than from the underlying calculations performed by $h$.} $\tau=0$ leads to deterministic, consistent chat replies; however, a poor performance. $\tau=1$ has higher flexibility, and $\tau=0.7$ seems to be a good compromise, aligning with literature observations.

\begin{figure}
%\centering
\begin{subfigure}[t]{0.14\textwidth}
\begin{tikzpicture}
\begin{axis}[
        axis on top,% ----
        width=\textwidth,
        scale only axis,
        enlargelimits=false, 
        ytick={0,25, 50, 75, 100},     
        xtick={0,50,100,150},
        ylabel={Accuracy [\%]},
        xlabel={Levenshtein distance [ ]},
        xmin=50,
        xmax=170,
        ymin=0,
        ymax=120,
        y label style={at={(axis description cs:-0.2,.5)},anchor=south}, 
        ]        
	\addplot graphics[xmin=45,xmax=175,ymin=0,ymax=120] {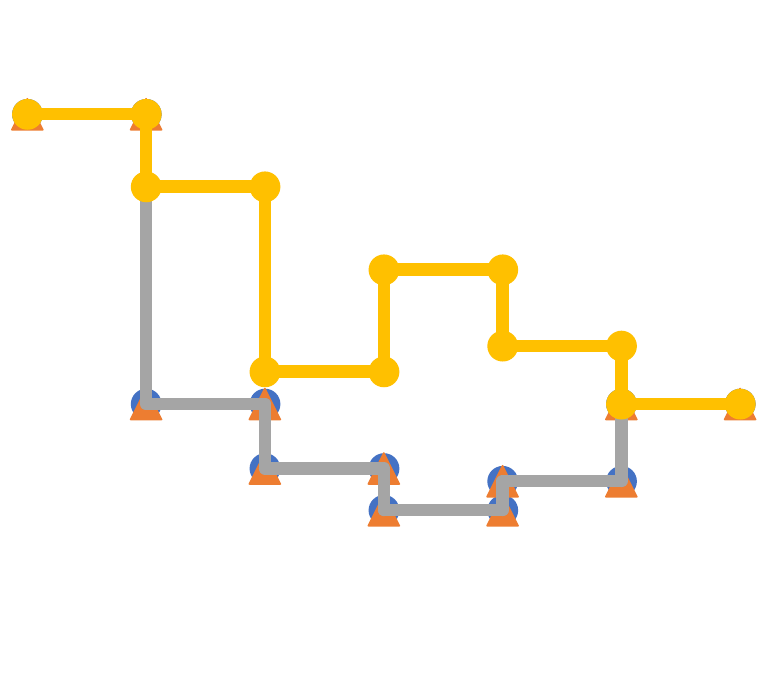};
  \end{axis}
\end{tikzpicture}
   \caption{Paraphrasing}
   \label{fig:csparaphrase}
\end{subfigure} \hspace{0.13\textwidth}
\begin{subfigure}[t]{0.14\textwidth} 
\begin{tikzpicture}
\begin{axis}[
        axis on top,% ----
        width=\textwidth,
        scale only axis,
        enlargelimits=false, 
        ytick={0,25, 50, 75, 100},     
        xtick={0, 2, 4, 6, 8, 10},
        ylabel={Accuracy [\%]},
        xlabel={Error rate [\%]},
        xmin=0,
        xmax=10,
        ymin=0,
        ymax=120,
        y label style={at={(axis description cs:-0.2,.5)},anchor=south},
        ]        
	\addplot graphics[xmin=-0.5,xmax=10.5,ymin=0,ymax=120] {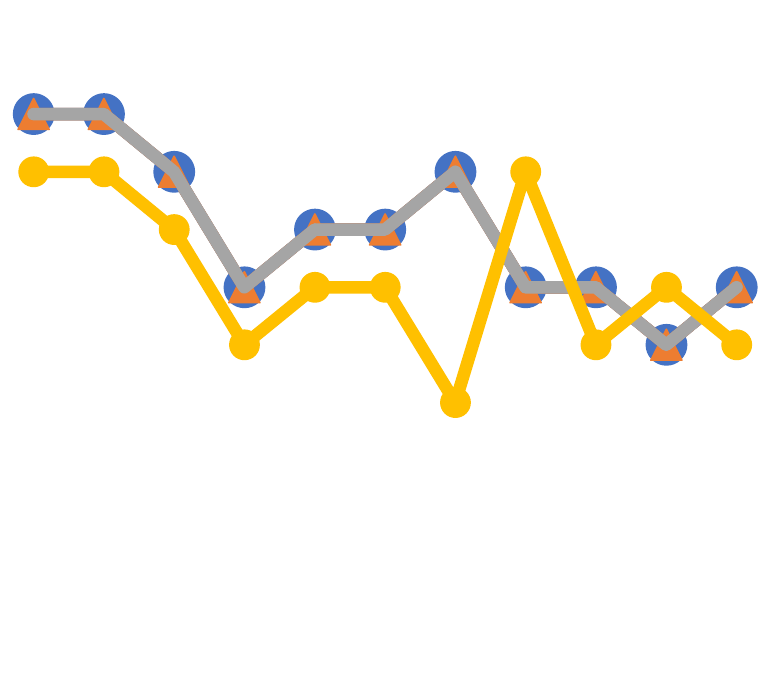};
  \end{axis}
\end{tikzpicture}
   \caption{Typographical}
   \label{fig:cstypos}
\end{subfigure}
\caption{Sensitivity of the user inputs $u_i$ when paraphrased (a) or includes typo errors (b); performance of $g$ to correctly identifying parameters $\alpha$ from text $u_i$ (\textcolor[RGB]{68,114,196}{\rule[0.15em]{0.8em}{0.3em}}), of the functional program $h$ to correctly output $\beta$ (\textcolor[RGB]{237,125,49}{\rule[0.15em]{0.8em}{0.3em}}), of $v$ to pass information from program to program (\textcolor[RGB]{165,165,165}{\rule[0.15em]{0.8em}{0.3em}}) and of $f$ to include parameters $\beta$ in response $r_i$ (\textcolor[RGB]{255,192,0}{\rule[0.15em]{0.8em}{0.3em}}).}
\label{fig:csuserinput}
\end{figure}
% \subsection{Negotating flexible contracts}
% This case study investigates whether the proposed system proposes alternatives back once a contract is found to be infeasible for the grid. This tests the flexibility of the approach.

% \subsection{Sensitivity to functions-naming}
% Importance of naming function in natural language... 
% examples... test by randomizing/anonymising strings
% Is tool used? Yes/no
% Are correct inputs to tool identified?

%\vspace{-1em}
\subsection{Sensitivity to user input \textcolor{black}{ in residential contracts}}
\textcolor{black}{Residential users specifically have heterogeneous knowledge about the energy system, varying intentions, and varying language fluency.} This study investigates the sensitivity of Alg. \ref{alg1} to user input $u_i$ \textcolor{black}{variations}, considering paraphrasing and typographical errors. The first part of this study uses a pre-trained model \cite{Pri21} to paraphrase user inputs. %TAnother model uses the pegasus architecture \cite{Zha20} that was fine-tuned for paraphrasing. (Memo JLC: Not used)
The baseline user input was $u_0=$\textit{'I use my washing machine $2$h and dishwasher $90$min every day. My TV is around 3h on and computer $4$h. The lights are on $14$h a day. My EV charges $1$h every day starting at 10:00.'}. 
The paraphrasor (with default settings) was applied $10$ times sentence by sentence on this user input. Then, $100$ combinations of the paraphrased sentences were uniformly sampled. The Levenshtein distance measures the difference between paraphrased and baseline sentences. For example, one of the four sentences in the baseline was \textit{'I use my washing machine $2$h and dishwasher $90$min every day.'} which could be paraphrased as \textit{'i use my washing machine $2$h and my dishwasher $90$min every day'} with a Levenshtein distance of $13$ or as \textit{'i am using my washing machine every day for $2$ hours and my dishwasher every $90$ minutes'} with a Levenshtein distance of $51$. As one can see, the latter paraphrasing led to a different meaning. Fig. \ref{fig:csparaphrase} presents the average accuracies within bins of width $20$.

The second part of this study investigates typographical errors. Different levels of error rates $\{0.01,0.02,\dots 0.1\}$ for typographical errors were considered, e.g., a rate of $3$\% means that there are $3$ typos in a text of $100$ characters. The typos were randomly sampled from character swaps, missing characters, adding extra characters, replacing a random character with keyboard-neighbour letters, replacing visually similar characters, skipping characters, adding or skipping a random space, repeating a character, or removing a letter in syntactic duplications. The Python code \cite{Kum21} simulated $10$ samples for each error rate, so $100$ different user inputs. For example, \textit{'I use my washing machine $2$h and xishwasher $9$ $0$min every day.'} has two typos and an error rate of $2$\% and \textit{'Iusc my washing acbime $2$h and dishwasher $90$min evvery day.'} has an error rate of $10$\%. \textcolor{black}{Together with the first part, the paraphrasing study in Fig. \ref{fig:csparaphrase}, this provides a comparison between unaltered inputs and controlled user-input variations. These settings make explicit how incorrect or distorted entries in $u_i$ propagate through the modelling chain and influence the correctness of $\beta$.} Fig. \ref{fig:cstypos} shows decreasing performance with increasing typos. An error rate above $1$-$2$\% may significantly influence the success of user engagement. \textcolor{black}{The system's strengths are the flexibility to heterogeneous user input. However, a minimal accuracy of the user input is needed: the intent needs to be correctly stated, as demonstrated in the first part, and the level of fluency needs to be high enough to extract the correct intents, as demonstrated in the second part.}

%vector of words are still very close
\begin{figure}
\centering
\includegraphics[width=0.22\textwidth]{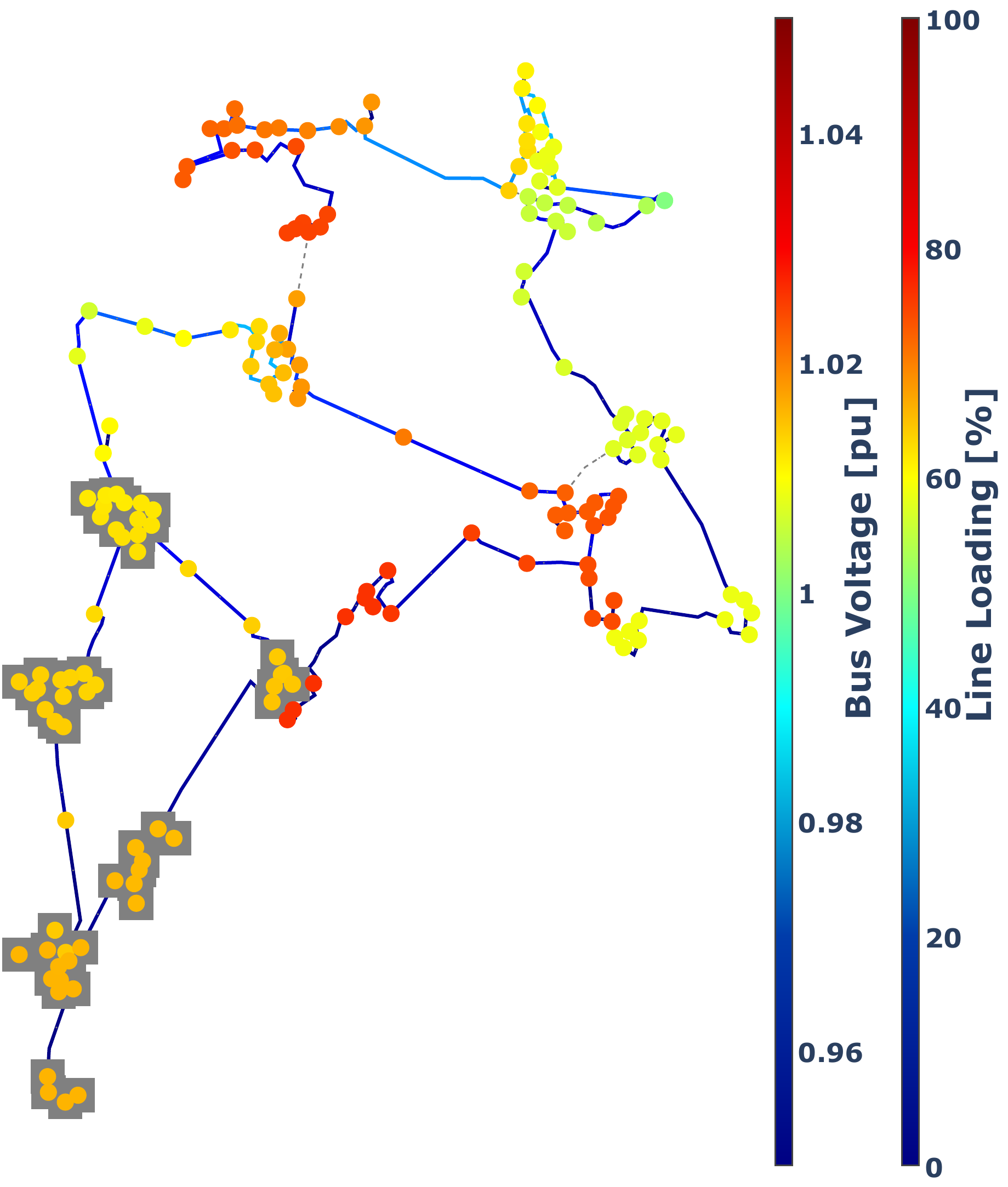}
\caption{Congested Oberrhein network. Locations are marked with a grey box where adding $2$MW constant power is physically infeasible at $19$h.} %\vspace{-2.5em}}
    \label{fig:usecaseMV}
\end{figure}

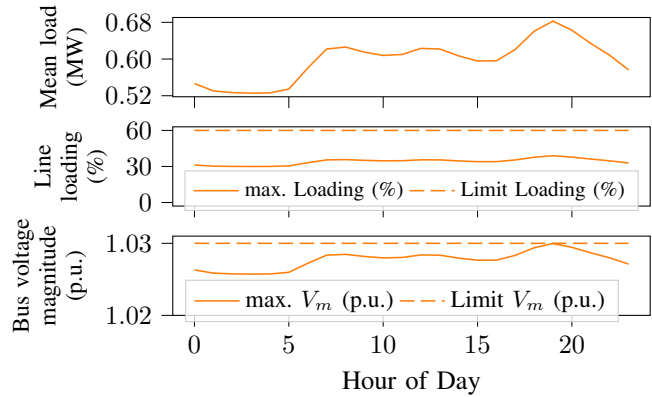
\begin{figure}%{0.41\textwidth}
\begin{center}
% This file was created with tikzplotlib v0.10.1.
\begin{tikzpicture}

\definecolor{darkgray176}{RGB}{176,176,176}
\definecolor{darkorange25512714}{RGB}{255,127,14}
\definecolor{lightgray204}{RGB}{204,204,204}
\definecolor{steelblue31119180}{RGB}{31,119,180}

\begin{groupplot}[group style={
  group size=1 by 3,
  vertical sep=0.4cm  % or whatever spacing you prefer
}]
\nextgroupplot[
tick align=outside,
tick pos=left,
height = 3em,
scale only axis,
width=0.35\textwidth,
x grid style={darkgray176},
xmin=-1.15, xmax=24.15,
xtick style={color=black},
xtick={-5,0,5,10,15,20,25},
% xticklabels={
%   $\mathrm{−5}$,
%   $\mathrm{0}$,
%   $\mathrm{5}$,
%   $\mathrm{10}$,
%   $\mathrm{15}$,
%   $\mathrm{20}$,
%   $\mathrm{25}$
% },
%xtick=\empty,
xticklabels={},
xlabel={},
y grid style={darkgray176},
ylabel={\small \parbox{2cm}{\centering Mean load\\(MW)}},
ymin=0.517776416505442, ymax=0.690563680134013,
ytick style={color=black},
ytick={0.52,0.6,0.68},
yticklabels={
  $\mathrm{0.52}$,
 % $\mathrm{0.56}$,
  $\mathrm{0.60}$,
%  $\mathrm{0.64}$,
  $\mathrm{0.68}$,
}
]
\addplot [semithick, darkorange25512714]
table {%
0 0.546567341891156
1 0.530521152462584
2 0.526740917319728
3 0.525630383034014
4 0.526291401319727
5 0.534486378462585
6 0.580167042748299
7 0.621750218462584
8 0.6260857207483
9 0.615194236748298
10 0.607818757319727
11 0.610039236748299
12 0.623124983319729
13 0.621829458176872
14 0.606893508462585
15 0.595764010748298
16 0.596187015319727
17 0.620957232176871
18 0.660637476462584
19 0.682709713605442
20 0.662805522176871
21 0.633990545034014
22 0.608295079319728
23 0.576281056462585
};

\nextgroupplot[
legend cell align={left},
legend columns=2,
legend style={
font=\footnotesize,
  fill opacity=0.8,
  draw opacity=1,
  text opacity=1,
  at={(0.5,0.45)},
  anchor=north,
  draw=lightgray204
},
tick align=outside,
tick pos=left,
height = 3em,
scale only axis,
width=0.35\textwidth,
x grid style={darkgray176},
xmin=-1.15, xmax=24.15,
xtick style={color=black},
xtick={-5,0,5,10,15,20,25},
% xticklabels={
%   $\mathrm{−5}$,
%   $\mathrm{0}$,
%   $\mathrm{5}$,
%   $\mathrm{10}$,
%   $\mathrm{15}$,
%   $\mathrm{20}$,
%   $\mathrm{25}$
% },
%xtick=\empty,
xticklabels={},
xlabel={},
y grid style={darkgray176},
ylabel={\small \parbox{2cm}{\centering Line \\loading\\(\%)}},
ymin=-2.95755550587203, ymax=62.997978833613,
ytick style={color=black},
ytick={0,30,60},
yticklabels={
  $\mathrm{0}$,
  $\mathrm{30}$,
  $\mathrm{60}$,
}
]
% \addplot [semithick, steelblue31119180]
% table {%
% 0 0.0404314113692949
% 1 0.0404252286859474
% 2 0.0404237600212512
% 3 0.0404233277409237
% 4 0.0404235851407356
% 5 0.040426764296634
% 6 0.0404440890770342
% 7 0.0404592769299001
% 8 0.0404608285319613
% 9 0.0404569192619269
% 10 0.0404542504042183
% 11 0.04045505572077
% 12 0.0404597696842798
% 13 0.0404593054065889
% 14 0.0404539144091621
% 15 0.0404498507630597
% 16 0.0404500059766981
% 17 0.0404589925375728
% 18 0.0404729788442631
% 19 0.0404805408397987
% 20 0.0404737285106969
% 21 0.0404636419733694
% 22 0.0404544232266246
% 23 0.0404426414439478
% };
% \addlegendentry{min. Loading (\%)}
\addplot [semithick, darkorange25512714]
table {%
0 31.1753441490035
1 30.2588780379683
2 30.0429074031252
3 29.9794562303071
4 30.01722423728
5 30.4853909973917
6 33.0928506292574
7 35.462972001539
8 35.709885718093
9 35.0895267378668
10 34.6692982947759
11 34.7958246506741
12 35.541271118166
13 35.4674851481681
14 34.6165732977398
15 33.9822303166479
16 34.0063445628204
17 35.4178060666598
18 37.6762974422942
19 38.931186592459
20 37.7996036889292
21 36.1599803420174
22 34.6964406864928
23 32.8711887619633
};
\addlegendentry{max. Loading (\%)}
\addplot [semithick, darkorange25512714, dash pattern=on 5.55pt off 2.4pt]
table {%
0 60
1 60
2 60
3 60
4 60
5 60
6 60
7 60
8 60
9 60
10 60
11 60
12 60
13 60
14 60
15 60
16 60
17 60
18 60
19 60
20 60
21 60
22 60
23 60
};
\addlegendentry{Limit Loading (\%)}

\nextgroupplot[
legend cell align={left},
legend columns=2,
legend style={
font=\small,
  fill opacity=0.8,
  draw opacity=1,
  text opacity=1,
  at={(0.48,0.45)},
  anchor=north,
  draw=lightgray204
},
tick align=outside,
tick pos=left,
height = 3em,
scale only axis,
width=0.35\textwidth,
x grid style={darkgray176},
xlabel={Hour of Day},
xmin=-1.15, xmax=24.15,
xtick style={color=black},
xtick={-5,0,5,10,15,20,25},
xticklabels={
$\mathrm{−5}$,
$\mathrm{0}$,
$\mathrm{5}$,
$\mathrm{10}$,
$\mathrm{15}$,
$\mathrm{20}$,
$\mathrm{25}$
},
y grid style={darkgray176},
ylabel={\small \parbox{2cm}{\centering Bus voltage \\magnitude \\(p.u.)}},
ymin=1.02, ymax=1.031,
ytick style={color=black},
ytick={0.98,1,1.02,1.03,1.04},
yticklabels={
  $\mathrm{0.98}$,
  $\mathrm{1.00}$,
  $\mathrm{1.02}$,
  $\mathrm{1.03}$,
  $\mathrm{1.04}$
}
]
% \addplot [semithick, steelblue31119180]
% table {%
% 0 1
% 1 1
% 2 1
% 3 1
% 4 1
% 5 1
% 6 1
% 7 1
% 8 1
% 9 1
% 10 1
% 11 1
% 12 1
% 13 1
% 14 1
% 15 1
% 16 1
% 17 1
% 18 1
% 19 1
% 20 1
% 21 1
% 22 1
% 23 1
% };
% \addlegendentry{min. $V_m$ (p.u.)}
\addplot [semithick, darkorange25512714]
table {%
0 1.02630835488615
1 1.02586506534392
2 1.02576027251897
3 1.02572946096748
4 1.0257478022189
5 1.02597483848257
6 1.02722856764574
7 1.0283524794082
8 1.02846871078588
9 1.02817637881887
10 1.02797777641304
11 1.02803762266271
12 1.02838935508726
13 1.02835460531008
14 1.02795282528145
15 1.02765205695298
16 1.02766351001842
17 1.02833120066679
18 1.02938863470967
19 1.02997037564079
20 1.02944598034337
21 1.02868017291099
22 1.02799061818724
23 1.0271226932087
};
\addlegendentry{max. $V_m$ (p.u.)}
% \addplot [semithick, steelblue31119180, dash pattern=on 5.55pt off 2.4pt]
% table {%
% 0 0.975
% 1 0.975
% 2 0.975
% 3 0.975
% 4 0.975
% 5 0.975
% 6 0.975
% 7 0.975
% 8 0.975
% 9 0.975
% 10 0.975
% 11 0.975
% 12 0.975
% 13 0.975
% 14 0.975
% 15 0.975
% 16 0.975
% 17 0.975
% 18 0.975
% 19 0.975
% 20 0.975
% 21 0.975
% 22 0.975
% 23 0.975
% };
% \addlegendentry{Limit $V_m$ (p.u.)}
\addplot [semithick, darkorange25512714, dash pattern=on 5.55pt off 2.4pt]
table {%
0 1.03
1 1.03
2 1.03
3 1.03
4 1.03
5 1.03
6 1.03
7 1.03
8 1.03
9 1.03
10 1.03
11 1.03
12 1.03
13 1.03
14 1.03
15 1.03
16 1.03
17 1.03
18 1.03
19 1.03
20 1.03
21 1.03
22 1.03
23 1.03
};
\addlegendentry{Limit $V_m$ (p.u.)}
\end{groupplot}

\end{tikzpicture}
   \caption{Various metrics over a typical $24$h day in the Oberrhein network. The max voltage over all buses $m$ peaks at $19$h due to a peak in the load nearly surpassing the voltage limits. Adding more load to specific buses may lead to network instabilities and require careful assessment.}
   \label{fig:pfoberrhein}
   \end{center}
\end{figure}

%\vspace{-1.75em}
\subsection{Autonomous connecting industry to medium voltage grids}

\begin{figure}[ht]
\scriptsize
%\captionsetup{format=plain}
\begin{userbubble}
\textbf{Developer:} Can you help me plan where I can build a new site of $2$MW for commercial use?
\end{userbubble}

\begin{assistantbubble}
\textbf{DSO:} Certainly, I can help you plan an electricity contract for your new site. To start, could you please provide the zip code for the location where you are considering building the site? Keep in mind the zip code should be between 0 and 146. Once we have the location, we'll proceed with assessing the power profile over a regular day. This will help us determine the feasibility of connecting to the grid at that location.
\end{assistantbubble}

\begin{userbubble}
\textbf{Developer:} Zip code is $5$
\end{userbubble}

\begin{assistantbubble}
\textbf{DSO:} Great! Now that we have the zip code, please provide the expected electricity usage in megawatt-hours (MWh) for each hour of a regular day at your site. This will be a 24-hour profile, with values typically between 0 and 4 MW for each hour. For example: 
\begin{itemize}
\item Hour 1: $1.5$ MW
\item Hour 2: $2.0$ MW
\item ...
\item Hour 24: $1.8$ MW
\end{itemize}
Please confirm the power profile for your site so we can assess its feasibility.
\end{assistantbubble}

\begin{userbubble}
\textbf{Developer:}  Could you help me make an estimate of the electricity demand profile? I plan to have regular opening hours for the commercial building with some shops opening longer than others. I consider to run several restaurants in the evening and open the shops until around $8$pm.
\end{userbubble}

\begin{assistantbubble}
\textbf{DSO:} 
Certainly! Let's create a rough estimate of your electricity demand profile based on typical usage patterns for commercial buildings with shops and restaurants. Here's a potential profile: 
\begin{itemize}
\item Hours $1-6$ (Midnight to $6$ AM): Minimum activity, primarily for basic lighting and security systems. Let's estimate around $0.5$ MW.
\item Hours $7-9$ ($7$ AM to $9$ AM): Shops start opening, so there will be an increase in demand. Estimate around $1.0$ MW.
\item Hours $10-13$ ($10$ AM to $1$ PM): Peak shopping hours, high activity. Estimate around $2.0$ MW.
\item Hours $14-17$ ($2$ PM to $5$ PM): Steady shopping activity. Estimate around $1.5$ MW.
\item Hours $18-20$ ($6$ PM to $8$ PM): Shops closing, restaurants starting to pick up, so demand remains high. Estimate around $2.0$ MW.
\item Hours $21-23$ ($9$ PM to $11$ PM): Peak restaurant hours, shops closed. Estimate around $2.5$ MW.
\item Hour $24$ ($11$ PM to Midnight): Late evening, restaurants closing. Estimate around $1.0$ MW.
\end{itemize}
Would you like to adjust any of these estimates, or should I proceed with this profile to assess its feasibility?
\end{assistantbubble}

\begin{userbubble}
\textbf{Developer:} Sounds reasonable. Thanks.
\end{userbubble}

\begin{assistantbubble}
\textbf{DSO:} \vspace{-1em}
\begin{center}
\begin{tcolorbox}[
    colback=green!10,
    colframe=green!60!black,
    arc=1mm,
    boxrule=0.8pt,
    left=2mm, right=2mm,
    top=0mm, bottom=0mm,
    enhanced,
    width=4cm,
    halign=center,
    valign=center,
    box align=base,
    boxsep=0.5mm,      % tight internal padding
    before skip=1pt,
    after skip=1pt,
    fontupper={\scriptsize\setlength{\baselineskip}{10pt}}  % smaller text and tighter line height
]
\tikzwrench~Used MVContractPlanne~\tikzwrench
\end{tcolorbox}
\end{center}
The analysis shows that the requested power at hour $20$ ($8$ PM to $9$ PM) is not feasible for connection to the grid at the specified location. We need to adjust the power demand for that specific hour. Would you like to reduce the power demand for hour $20$, or make adjustments elsewhere in the profile to accommodate the required power? Let me know how you'd like to proceed, and we can reassess the feasibility.
\end{assistantbubble}

\begin{userbubble}
\textbf{Developer:} Yes, what if I close already the shops earlier? Let's say at $6$pm. Can you check if this resolves the issue?
\end{userbubble}

\begin{assistantbubble}
\textbf{DSO:} Sure! If you close the shops earlier at 6 PM, we can adjust the power profile as follows: 
\begin{itemize}
\item Hours $18-19$ ($6$ PM to $7$ PM): Reduce demand as shops start closing, estimate around $1.5$ MW instead of $2.0$ MW.
\item Hour $20$ ($8$ PM to $9$ PM): Restaurants remain open, we'll reduce slightly to $1.8$ MW to see if it's feasible.
\end{itemize}
Let's reassess the profile with these adjustments. 
\begin{center}
\begin{tcolorbox}[
    colback=green!10,
    colframe=green!60!black,
    arc=1mm,
    boxrule=0.8pt,
    left=2mm, right=2mm,
    top=0mm, bottom=0mm,
    enhanced,
    width=4cm,
    halign=center,
    valign=center,
    box align=base,
    boxsep=0.5mm,      % tight internal padding
    before skip=1pt,
    after skip=1pt,
    fontupper={\scriptsize\setlength{\baselineskip}{10pt}}  % smaller text and tighter line height
]
\tikzwrench~Used MVContractPlanne~\tikzwrench
\end{tcolorbox}
\end{center}
Great news! With the adjusted power profile, all times are now feasible for connection to the grid at the specified location. We can proceed with signing an electricity contract based on this profile. If you have any further questions or need additional adjustments, feel free to let me know!
\end{assistantbubble}
\caption{Chat between developer and DSO with WolframAlpha LLM. The function (green) runs AC power flows on the requested profile (Sec. \ref{sec:MVgrid}). }
\label{fig:examplechat}
\end{figure}

\textcolor{black}{The high complexity of customisation connection requests leads to long processing times.}
This case study investigates whether the approach \textcolor{black}{shows potential to reduce complexity and accelerate the} planning in distribution grids, bridging commercial business developers and a DSO. We consider the Oberrhein MV grid with a high feed-in scenario and an increased load that varies over time. The DSO has voltage issues in specific areas at a specific time of the day (at $19$h see Fig. \ref{fig:pfoberrhein}). A developer considers building a large commercial building with a nominal power rating of $2$MW, and can decide between multiple locations. The use of the commercial building determines a daily electricity profile; the developer of the project has flexibility in renting parts of the building to various commercial sectors, e.g. adjusting opening times depending on power system constraints. Some site locations don't require compromising building plans; others do require compromises. Deciding this is complex, as the best business case for this commercial building requires assessing location and possible use cases simultaneously. The developer can use the chat to assess whether the intended location and power profile are feasible, e.g. nominal (constant) power is physically infeasible at the locations marked in Fig. \ref{fig:usecaseMV}. The state-of-the-art planning of this process would involve teams from multiple stakeholders, such as developers, DSO and others, which takes significant planning time. The proposed chat, however, iterates with the developer until a feasible combination of use case and location, and grid requirements is found in near-real time. Fig. \ref{fig:examplechat} is an example chat. The MVContractPlanner (green box in figure) is the function $h(\alpha)$ that runs Alg. \ref{alg2}, calling the AC power flow for all $24h$ on the Oberrhein system on bus $b=5$ as introduced in Sec. \ref{sec:MVgrid}. The bus $b=5$ and the power values $P_t$ are extracted from the chat and passed as input $\alpha=\{b, P_t\}$ to this function $h(\alpha)$. This study demonstrated \textcolor{black}{the approach is promising to reduce complexity through} automating a connection request for a DSO, interfacing customer management and network analysis tooling. Note that the state-of-the-art connection request may take around months to be completed; here, these can be resolved within minutes with possibly minimal human oversight from a DSO.

%\vspace{-1em}
\subsection{\textcolor{black}{Accelerate} outage planning in transmission system}
\begin{figure}[ht]
\begin{center}
    \includegraphics[width=0.33\textwidth]{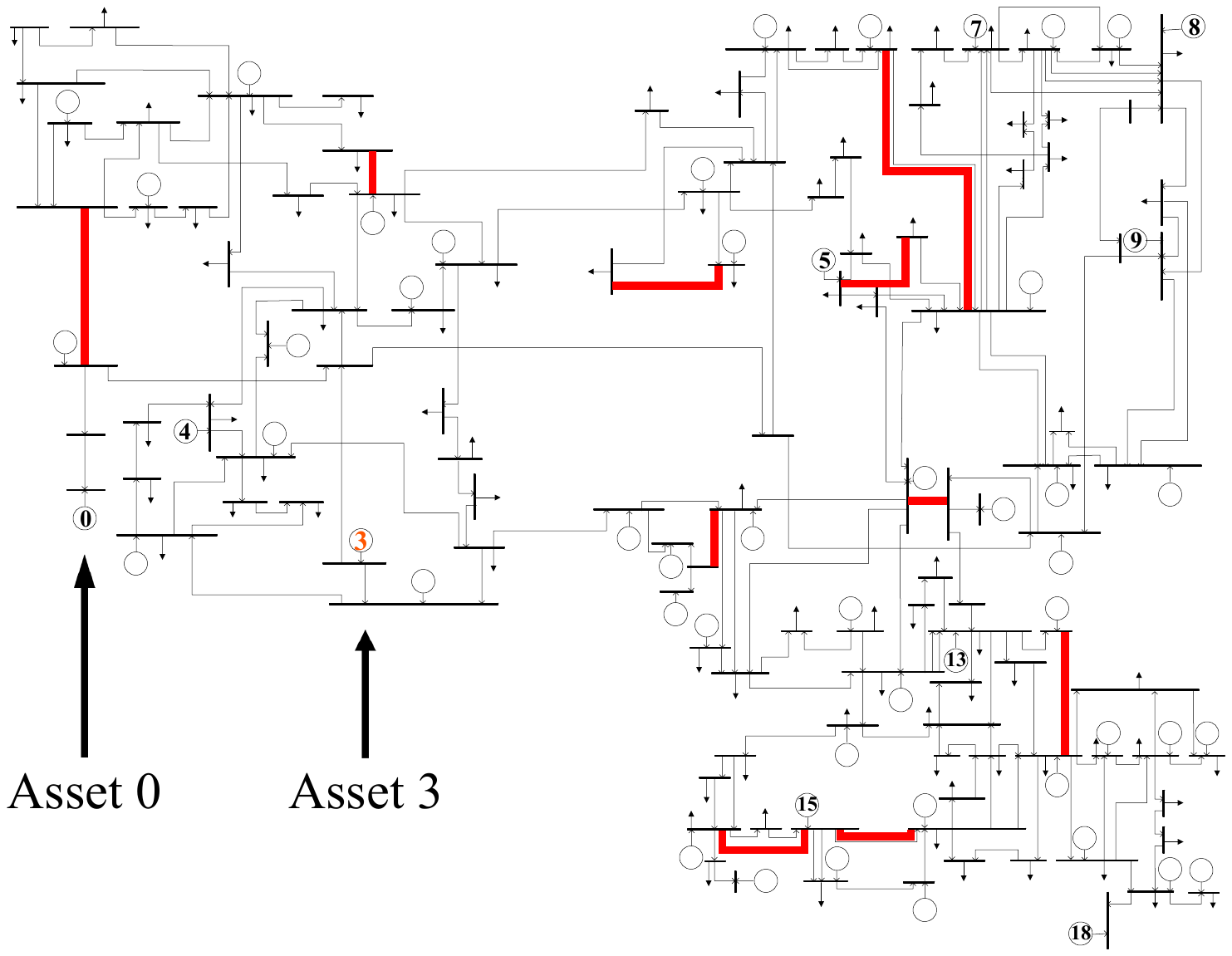}
   \caption{IEEE $118$-bus system where scheduling generator $3$ at the same time as generator $0$ would lead to N-$1$ insecurity of the red marked lines. Other times shown in Fig. \ref{fig:gantt} do not lead to insecurity. }
   %\vspace{-1em}
   \label{fig:118bus}
   \end{center}
\end{figure}

\begin{figure}
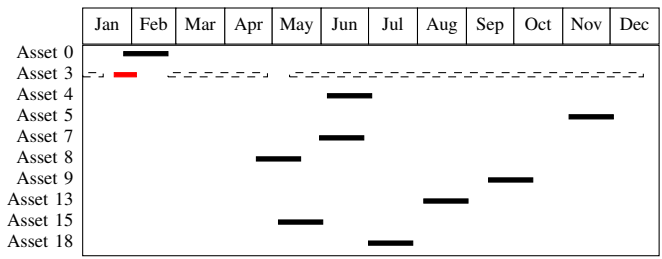

\centering
\begin{ganttchart}[
    x unit=0.021cm,
    y unit chart=0.28cm,         % bar row height
    y unit title=0.8cm,         % compress title row height
    %title height=0.2,              % allow full label box
    title top shift=-0.62,        % pull label upward inside row
    bar height=0.2,
    bar/.append style={fill=black},
    bar label font=\scriptsize,
    title label font=\scriptsize,
    bar label anchor/.style={text width=2.5cm, align=left},
    time slot unit=day,
    hgrid=false,
    vgrid=false
  ]{1}{365}

  % Abbreviated months with small font, no grid background
  \gantttitle{Jan}{31}
  \gantttitle{Feb}{28}
  \gantttitle{Mar}{31}
  \gantttitle{Apr}{30}
  \gantttitle{May}{31}
  \gantttitle{Jun}{30}
  \gantttitle{Jul}{31}
  \gantttitle{Aug}{31}
  \gantttitle{Sep}{30}
  \gantttitle{Oct}{31}
  \gantttitle{Nov}{30}
  \gantttitle{Dec}{31} 
  \ganttbar{Asset 0}{27}{54} \\
  \ganttbar[bar/.append style={fill=none, draw=black, dashed}]{Asset 3}{1}{13}
  \ganttbar[bar/.append style={fill=red,draw=red}]{}{21}{34}
  \ganttbar[bar/.append style={fill=none, draw=black, dashed}]{}{55}{117}
  \ganttbar[bar/.append style={fill=none, draw=black, dashed}]{}{132}{355} \\
  \ganttbar{Asset 4}{156}{183} \\
  \ganttbar{Asset 5}{309}{336} \\
  \ganttbar{Asset 7}{151}{178} \\
  \ganttbar{Asset 8}{111}{138} \\
  \ganttbar{Asset 9}{258}{285} \\
  \ganttbar{Asset 13}{217}{244} \\
  \ganttbar{Asset 15}{125}{152} \\
  \ganttbar{Asset 18}{182}{209}
\end{ganttchart}
\caption{Gantt chart of scheduled outages for $10$ generators. The initially requested time (red) of the owner asset $3$ is not physically feasible. The proposed chat provides alternative times back to the owner (dashed lines).\vspace{1em}}
\label{fig:gantt}
\end{figure}

%\vspace{-0.25em}
\textcolor{black}{Current outage planning takes currently months to complete, and the requests pile up due to complex planning constraints and frequent follow-up discussions. This study investigates the potential of the proposed approach to accelerate and} customise requests for disconnecting assets from the grid. An asset owner plans maintenance of the generator $3$ in the IEEE $118$-bus system, Fig. \ref{fig:118bus}. The asset owner interacts with the grid network studies (calling Alg. \ref{alg3}) in near real-time via the chat, Fig. \ref{fig:transmissionchat}. The initial request to maintain in week $3$ is assessed with OutagePlanner $h(\alpha)$ running Alg. \ref{alg2}. The OutagePlanner solved several SCOPFs showing the request would lead to N-$1$ insecurity. Another generator $0$ is already scheduled for maintenance during these times (Fig. \ref{fig:gantt}). These two generators, $0$ and $3$, are close to each other (Fig. \ref{fig:118bus}), hence simultaneous maintenance is not N-$1$ secure. The chat proposes alternative times (Fig. \ref{fig:transmissionchat}) where the week $8$ suits the owner, and the scheduled maintenance is confirmed. \textcolor{black}{The time for the proposed LLM-based system and GUROBI to communicate with the user and solve $365$ SCOPFs took around $1$ min. In comparison, the state-of-the-art approach requires requests and often follow-up conversations between a TSO with the asset owner, which overall can take several months.} An additional benefit is that grid code documentations, which change frequently, can be instantly updated without reskilling employees; here, updating a document to the chat context $\omega_i$. %This fast updating is an additional benefit as these documents frequently change.

\begin{figure}
\scriptsize
\begin{userbubble}
\textbf{Asset owner:} Hello. I would like to plan an outage of my generator asset. Can you help me with that?
\end{userbubble}

\begin{assistantbubble}
\textbf{TSO:} Of course! To help you plan the outage for your generator asset, I'll need the following information:
\begin{itemize}
\item 1. \textbf{Starting Week}: A number between $1$ and $50$ indicating when you want the outage to begin.
\item 2. \textbf{Duration}: The duration of the outage in weeks, between $1$ and $10$.
\item 3. \textbf{Asset Number}: The asset number of the generator you wish to plan the outage for, which should be a number between $0$ and $18$.
\end{itemize}
Once I have this information, we can proceed with planning the outage.
\end{assistantbubble}

\begin{userbubble}
\textbf{Asset owner:} The asset number is $3$. I plan to take it in outage at week $3$ of this year. The duration is $2$ weeks.
\end{userbubble}

\begin{assistantbubble}
\textbf{TSO:}  %\vspace{-1em}
\begin{center}
\begin{tcolorbox}[
    colback=green!10,
    colframe=green!60!black,
    arc=1mm,
    boxrule=0.8pt,
    left=2mm, right=2mm,
    top=0mm, bottom=0mm,
    enhanced,
    width=4cm,
    halign=center,
    valign=center,
    box align=base,
    boxsep=0.5mm,      % tight internal padding
    before skip=1pt,
    after skip=1pt,
    fontupper={\scriptsize\setlength{\baselineskip}{10pt}}  % smaller text and tighter line height
]
\tikzwrench~Used OutagePlanner~\tikzwrench
\end{tcolorbox}
\end{center}
It seems that starting the outage for your generator asset number $3$ in week $3$ is not possible. However, there are some alternatives you can consider:
\begin{itemize}
\item 1. \textbf{Start earlier}: You could begin the outage for a duration of $2$ weeks starting between day $1$ and day $13$.
\item 2. \textbf{Start later}: You could also choose to start the outage for a duration of $2$ weeks between day $55$ and day $117$.
\end{itemize}
Let me know if any of these options work for you, or if you would like to explore other possibilities.
\end{assistantbubble}

\begin{userbubble}
\textbf{Asset owner:} Starting maintenance at day $56$ would work for us. Can you confirm this works? 
\end{userbubble}

\begin{assistantbubble}
\textbf{TSO:}  %\vspace{-1em}
\begin{center}
\begin{tcolorbox}[
    colback=green!10,
    colframe=green!60!black,
    arc=1mm,
    boxrule=0.8pt,
    left=2mm, right=2mm,
    top=0mm, bottom=0mm,
    enhanced,
    width=4cm,
    halign=center,
    valign=center,
    box align=base,
    boxsep=0.5mm,      % tight internal padding
    before skip=1pt,
    after skip=1pt,
    fontupper={\scriptsize\setlength{\baselineskip}{10pt}}  % smaller text and tighter line height
]
\tikzwrench~Used OutagePlanner~\tikzwrench
\end{tcolorbox}
\end{center}
Starting the maintenance for your generator asset number 3 at day 56, which corresponds to week 8, is confirmed to be possible. You can proceed with the outage for the specified duration of 2 weeks.

If you have any more questions or need further assistance, feel free to ask!
\end{assistantbubble}
\caption{Chat to plan an outage in the transmission IEEE $118$-bus system using WolframAlpha LLM. The green functions perform SCOPFs. \vspace{-1em}}
\label{fig:transmissionchat}
\end{figure}

\subsection{Privacy in transmission assets maintenance planning}
\label{sec:cspriv}
\begin{figure}
\centering
\begin{tikzpicture}
\begin{axis}[
        axis on top,% ----
        width=0.15\textwidth,
        scale only axis,
        enlargelimits=false, 
        ytick={1,2,3,4,5,6},  
        yticklabels= {1,2,3,4,5,$n$},  
        xtick={0,1,2,3,4,5,6,7,8,9},
        ylabel={Asset number [ ]},
        xlabel={Time horizon [weeks]},
        xmin=-0.5,
        xmax=9.5,
        ymin=0.5,
        ymax=6.8,
        ]        
	\addplot graphics[xmin=-0.5,xmax=9.5,ymin=0.5,ymax=6.8] {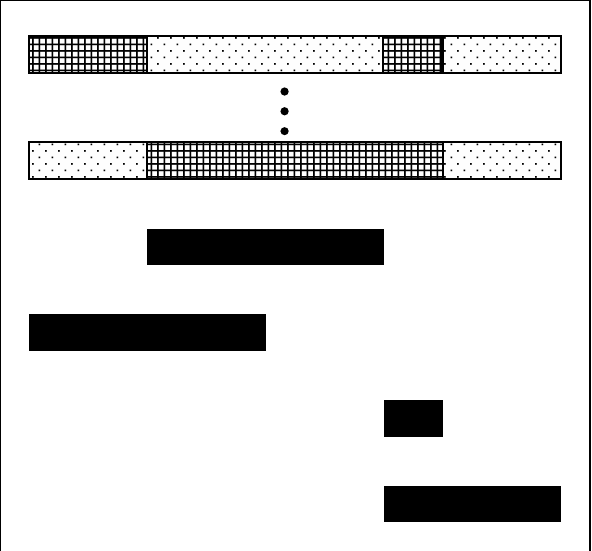};
  \end{axis}
\end{tikzpicture}
\caption{Possible security implications of the maintenance planning approach (see Fig. \ref{fig:approachSCOPF}). A brute force extruder could vary $\alpha$ to learn about the power network; however, the extruder would maximally obtain information about the maintenance possibilities of each generator through $\beta$, for example, when assets $5$ to $n$ can be switched off.\vspace{0.5em}}
\label{fig:maxinformation}
\end{figure}

\textcolor{black}{LLM-based systems have limitations with security and privacy of the user inputs. } This analysis investigates whether the proposed approach may have security breaches that could make sensitive information \textcolor{black}{from a TSO} available to the user\textcolor{black}{ of the chat}. \textcolor{black}{In a real-world implementation,} the information within the Python scripts and grid data \textcolor{black}{would} not be publicly accessible. However, an attacker may investigate many variations of inputs $\alpha$, trying to learn grid behaviour and settings, like existing planned outages, for example, which can be considered sensitive information. Fig. \ref{fig:maxinformation} shows the maximal information one could learn. One could learn the times for each generator when it's feasible to switch it off and when not, as shown with black bars through many requests $\alpha$ and analysing $\beta$. Analysing these may reveal information when (and where) the grid is more congested and weak. One could limit use only for internal stakeholders or limit the number of requests to prevent these attacks. \textcolor{black}{In the two, the normal and adversarial querying settings, the chat-user only receives the feasibility outcomes $\beta$ produced by the functional programs $h$, while no internal grid data or intermediate quantities become part of the chat context $\omega_i$. This structural separation ensures that the information exposed through $\alpha$ and $\beta$ is limited by the tool interface itself. We discuss the limitations on privacy and security further in Sec. \ref{sec:disc}. } %A more comprehensive analysis on this subject is beyond the scope of this paper.

%\vspace{-0.8em}

\vspace{-1em}
\subsection{\textcolor{black}{Discussion}}
\label{sec:disc}
%\vspace{-2em}
To the best of our knowledge, the proposed approach is the first to apply LLMs to electricity contract design, managing end-users of the grid, system operators, or energy suppliers. The key benefit of the LLM is to coach the user through the contract design, where the user has heterogeneous knowledge about this process. This makes the contract design scalable to a large number, where the state-of-the-art contracts have to be standardised and typically are supported with long guidelines, rules, and FAQs. This paper presented the initial idea and studied some algorithmic properties (like temperature), sensitivity to user input, privacy, and the impact this system can have within three power system use cases.

While the case studies present a promising impact, the methods and study have several limitations. LLMs have vulnerabilities that need to be studied in detail before deploying this system. The vulnerabilities typically involve security (e.g., prompt hacking and adversarial attacks) and privacy attacks \cite{Das25}, and various methods for system design can be applied (e.g., API hardening or paraphrasing). Sec. \ref{sec:cspriv} studied one privacy attack on grid data with a basic brute force strategy. Beyond the proposed measures to secure privacy, the hosting setup and data policies primarily determine privacy. For example, the LLM system can be either hosted in a cloud environment or in a private environment, where the data stays in this environment, ensuring high privacy. In either environment, regular chats are typically not used to retrain models live, ensuring the privacy of entered information. When self-hosted, the developer can specify the type of information recorded from the user, and the chat is not stored on the server. Beyond these future needs on privacy and security, this work has not yet explored fine-tuning, a wider variety of the intent of users, exhaustive robustness to very high error rates, real-world variety beyond typographical errors, or evolving model behaviours. No standard benchmarks exist yet to compare this model against others. This study also did not focus on a variety of baselines and improved evaluation frameworks that go beyond the accuracy of responses and parameter correctness to exhaustively compare different models and settings. The real-world applicability needs to be studied further, for example, to consider dynamic operating envelopes that can be defined through contracted flexibility, the expected cost reduction, and customer adoption of the system. While the methodological framework is independent of a specific interface, designing a dedicated User Interface for deployment is a natural next step. All the aforementioned limitations need to be addressed to develop this system further for production readiness. 

%https://blog.paperspace.com/measuring-text-similarity-using-levenshtein-distance/
%\subsection{Studying feasible grid contracts}
%This case study investigates whether the proposed system provides the correct results of the studies. Investigate the limits..., e.g. increase urgency.

%\subsection{Sensitivity to tool instructions}

%https://thepythoncode.com/article/paraphrase-text-using-transformers-in-python?utm_content=cmp-true

%\subsection{Discussion}
%https://medium.com/@2twitme/llm-beyond-its-core-capabilities-as-ai-assistants-or-agents-704ffb972934

\vspace{-0.5em}
\section{Conclusion} \label{sec:conclusion}
%\vspace{-0.25em}
\textcolor{black}{This paper introduced a new interaction layer between end-users and system operators that has not existed in current practice, where such studies are carried out manually and only by experts. Applying this layer via LLM-based} chats to customise contracts can drastically reduce the transaction costs of interacting with electricity users. Electrical grids conventionally standardise energy supply contracts, not considering the hidden flexibility in the diversity of preferences and devices. The reduction of transaction costs can make the customisation of supply contracts profitable, reduce outage planning durations, and enable new tooling for SMEs to directly test their electricity connections. For example, we demonstrate a $7$-fold increase of EVs that can connect to the grid, exploiting the heterogeneity of devices and preferences. The chat system shows low error rates, robustness to typos, and adaptability. We believe this approach can enhance the digitalisation of the electricity system and \textcolor{black}{enable} the energy transition. 

\vspace{-3em}
\textcolor{black}{
\section*{Acknowledgement} 
We thank our industrial collaborators in discussing these three use cases (Vikaran Khanna from National Grid ESO, UK, Luc Nies from Alliander, the Netherlands, and Arjan van Voorden, Stedin, the Netherlands). Some parts of the manuscript benefited from grammar and clarity improvements using automated tools, including Grammarly and ChatGPT. The authors remain responsible for all content.}

%\vspace{-1.5em}
\bibliographystyle{IEEEtran}
\bstctlcite{IEEEexample:BSTcontrol}
\bibliography{ref.bib}

\end{document}